# EXPERIMENTAL AND MODELING STUDY OF THE OXIDATION OF XYLENES

F. BATTIN-LECLERC, R. BOUNACEUR, N. BELMEKKI and P.A. GLAUDE

*Département de Chimie Physique des Réactions, UMR 7630 CNRS, INPL-ENSIC*

*1, rue Grandville, BP 451, 54001 NANCY Cedex - France*

**ABSTRACT**

This paper describes an experimental and modeling study of the oxidation of the three isomers of xylene (ortho-, meta- and para-xylenes). For each compound, ignition delay times of hydrocarbon-oxygen-argon mixtures with fuel equivalence ratios from 0.5 to 2 were measured behind reflected shock waves for temperatures from 1330 to 1800 K and pressures from 6.7 to 9 bar. The results show a similar reactivity for the three isomers.

A detailed kinetic mechanism has been proposed, which reproduces our experimental results, as well as some literature data obtained in a plug flow reactor at 1155 K showing a clear difference of reactivity between the three isomers of xylene. The main reaction paths have been determined by sensitivity and flux analyses and have allowed the differences of reactivity to be explained.

## INTRODUCTION

Aromatic compounds are present in significant amounts in gasolines (~35%) and diesel fuels (~30%) [1]. Nevertheless, detailed chemical kinetic models for the combustion and the oxidation of aromatic compounds are still scarce and mainly restricted to benzene [2-11] or toluene [12-24]. We have recently published a paper presenting an experimental and modeling study of the oxidation of toluene [24] that we have used as a basis for this work about the oxidation of xylenes.

Xylenes are important components of gasolines (up to 10 % for the sum of the three isomers), with high research octane numbers (117.5 for meta-xylene and 116.4 for para-xylene) close to that of toluene (120) [1]. The existing experimental studies concerning the reactions with oxygen of the three isomers of xylenes have been obtained in a flow reactor [25, 26] at ~1150 K, in a rapid compression machine [27, 28] between 600 and 900 K and in a single cylinder spark ignition engine [29]. The oxidation of meta and para-xylenes has also been studied in a jet-stirred reactor [30, 31] between 900 and 1300 K and that of para-xylene in a shock tube [30] between 1450 and 1760 K. Qualitative mechanisms explaining the experimental phenomena observed have been proposed [25, 28], but, up to now, only one detailed kinetic mechanism has been published in the case of para-xylene [30]. The purpose of this paper is to present new experimental data for the oxidation of ortho-, meta- and para-xylenes obtained in a shock tube from 1300 to 1820 K and pressures from 6.7 to 9 bar and to propose a mechanism able to model these experimental results, as well as literature data obtained in a flow reactor at ~1150 K [25, 26].

## EXPERIMENTAL IGNITION DELAYS TIMES

As it is already described in detail in previous papers [11, 33, 34], we will just recall here the main features of this experimental device. The reaction (400 cm long, 7.8 cm i.d.) and the driver (90 cm long, 12.8 cm i.d.) parts of this stainless steel shock tube were separated by two terphane diaphragms, which were ruptured by decreasing suddenly the pressure in the space separating them. The driver gas was helium. The incident and reflected shock velocities were measured by four piezo-electric pressure transducers located along the reaction section.

As in our previous work [11, 24, 33, 34], the temperature and the pressure of the test gas behind the incident and the reflected shock waves were derived from the value of the incident shock velocity by using ideal one-dimensional shock equations.

The onset of ignition was detected by OH radical emission at 306 nm through a quartz window with a photomultiplier fitted with a monochromator at the end of the reaction part. The last pressure transducer was located at the same place along the axis of the tube as the quartz window. The ignition delay time was defined as the time interval between the pressure rise measured by the last pressure transducer due to the arrival of the reflected shock wave and the rise of the optical signal by the photomultiplier up to 10% of its maximum value.

Oxygen (99.5% pure), argon and helium (both 99.995% pure) were supplied by Alphagaz - L'Air Liquide. Para-xylene (99 % pure) was provided by Merck-Schuchardt and ortho-xylene (98 % pure) and meta-xylene (97 % pure) by Sigma-Aldrich. Fresh reaction mixtures were manometrically prepared every day and mixed using a recirculation pump. Before each introduction of the reaction mixture (from 100 to 400 kPa), the reaction section was flushed with pure argon and evacuated (below $10^{-2}$ kPa), in order that the residual gas was mainly argon. The study was performed under the following experimental conditions (after the reflected shock):

- Temperature (T) from 1330 to 1800 K.
- Pressure (P) from 6.7 to 9 bar.
- For each of the three isomers, mixtures (xylene : oxygen : argon : in molar percent) were (0.375 : 7.875 : 91.75), (0.375 : 3.937 : 95.687), (0.375 : 1.969 : 97.653) and (0.625 : 6.562 : 92.812), respectively, which correspond to three different equivalence ratios ($\Phi$ =

0.5, 1 and 2) and to two different concentrations of xylene (0.375 % and 0.625 %) and lead to ignition delay times from 3 to 800 μs.

Tables I to III and Figures 1 to 3 present the experimental results thus obtained.

**TABLES I TO III – FIGURES 1 TO 3**

The scatter of the results is larger than in the cases of benzene [11] or toluene [24] ; that is probably due to the lower volatibility of xylenes (boiling points between 138 and 142 °C) compared to that of toluene (boiling point = 111°C) or benzene (boiling point = 80 °C) inducing then more condensation and absorption problems. For the twelve studied mixtures, ignition delay times (τ) decrease when temperature rises and varies exponentially vs. 1000/T. For each xylene, it is also shown that for a given T, ignition delay times increase with the equivalence ratio of the mixture for a given concentration of hydrocarbon, and slightly decrease with concentration. It is worth noting that in the case of toluene [24], ignition delay times notably increase when the concentration of hydrocarbon increases, while they decrease in the case of benzene [11].

**FIGURE 4**

Figure 4, shows that, under the same experimental conditions, the three isomers of xylenes have the samilar reactivity.

## DESCRIPTION OF THE REACTION MECHANISMS

These three mechanisms, which are available on request, have been written in the CHEMKIN II [32] format and include three parts:

- A primary mechanism containing 53 reactions for ortho-xylene and para-xylene and 51 reactions for meta-xylene, in which only xylene and oxygen are considered as molecular reactants,
- A secondary mechanism including 163 reactions for ortho-xylene, 137 reactions for meta-xylene and 138 reactions for para-xylene, in which the reactants are the main molecular products (apart toluene) formed by the primary mechanism. The primary and secondary mechanism for ortho-xylene includes then 216 reactions.

- The mechanisms for the oxidation of benzene [11] and toluene [24] connected to a $C_0$-$C_6$ reaction base, which is described in recent papers [33, 34].

As shown by Emdee et al. [25, 26], benzene and toluene are some important products of the oxidation of xylenes. As previously presented [11], the mechanism for the oxidation of benzene contains 135 reactions and includes the reactions of benzene and of cyclohexadienyl, phenyl, phenylperoxy, phenoxy, hydroxyphenoxy, cyclopentadienyl, cyclopentadienoxy and hydroxycyclopentadienyl free radicals, as well as the reactions of ortho-benzoquinone, phenol, cyclopentadiene, cyclopentadienone and vinylketene, which are the primary products yielded. The mechanism for the oxidation of toluene contains 193 reactions and includes the reactions of toluene and of benzyl, tolyl, peroxybenzyl (methylphenyl), alcoxybenzyl and cresoxy free radicals, as well as the reactions of benzaldehyde, benzyl hydroperoxyde, cresol, benzylalcohol, ethylbenzene, styrene and bibenzyl.

The $C_0$-$C_6$ reaction base was constructed from a review of the recent literature and is an extension of our previous $C_0$-$C_2$ reaction base [35, 33]. This $C_0$-$C_2$ reaction base includes the reactions of radicals or molecules including carbon, hydrogen and oxygen atoms and containing less than three carbon atoms. The kinetic data used in this base were taken from the literature and are mainly those proposed by Baulch *et al.* [36] and Tsang *et al.* [37]. The $C_0$-$C_6$ reaction base includes reactions involving $C_3H_2$, $C_3H_3$, $C_3H_4$ (allene and propyne), $C_3H_5$ (three isomers), $C_3H_6$, $C_4H_2$, $C_4H_3$ (2 isomers), $C_4H_4$, $C_4H_5$ (5 isomers), $C_4H_6$ (1,3-butadiene, 1,2-butadiene, methyl-cyclopropene, 1-butyne and 2-butyne), as well as the formation of benzene [33, 34]. Pressure-dependent rate constants follow the formalism proposed by Troe [38] and efficiency coefficients have been included. This reaction base was constructed in order to model experimental results obtained in a jet-stirred reactor for methane and ethane [35], profiles in laminar flames of methane, acetylene and 1,3-butadiene [33] and shock tube auto-ignition delays for acetylene, propyne, allene, 1,3-butadiene [33], 1-butyne and 2-butyne [34].

Heats of formation, specific heats, and entropies of the molecules or radicals considered have been mainly calculated using software THERGAS [39], based on the group and bond additivity methods proposed by Benson [40], and stored as 14 polynomial coefficients, according to the CHEMKIN II formalism [32]. It must be kept in mind that the precision obtained by using group additivity methods to estimate heats of formation is around 2 kcal/mol for molecules and 4 kcal/mol for radicals [40]. Table IV presents, in the case of ortho-xylene the names, the formulae and the heats of formation of the 35 aromatic species which are not included in the mechanism for the oxidation of benzene [11] and toluene [24]. The heats of formation of the isomers involved in the mechanisms are very close whatever the ortho, meta or para position of the ring substituents, apart from xylylenes. Ortho and para-xyxylenes have been considered in their quinoid structure, which is the most stable ; the formation of meta-xylylene which is in a triplet electronic ground state with a heat of formation, $\Delta H_f$ (298K) = 74 kcal/mol [41], has not been considered.

**TABLE IV**

The primary and secondary mechanisms for the oxidation of xylenes are presented in table V, as well as the references related to the kinetic data. For the shortness this table, a reaction common to the three isomers is written only once without specifying the type of isomer involved. The mechanism of a given isomer includes all the common reactions, as well as the reactions specific to this type of isomer. The rate parameters for the metatheses are not displayed as they are directly derived from the mechanism for toluene [24] or benzene [11].

**TABLE V**

Primary mechanism

The primary mechanism includes the reactions of xylene molecules and of methylbenzyl, methyltolyl, tolylmethoxyl and methylcresoxyl free radicals**.**

Unimolecular decompositions (1 and 2, numbers referring to Table V) of xylene molecules give either methylbenzyl radicals and H atoms or tolyl and methyl radicals. Reaction 1 is one of the few elementary steps involving xylenes which have been experimentally investigated. Da Costa *et al.* [42] have shown that the three isomers exhibit a very close reactivity for the decomposition giving H atoms and have proposed the rate expressions that we have used with a A-factor divided by two. For reaction 2, we have considered a rate constant four times higher than that used for the similar reaction in the case of toluene: the presence of two possible methyl groups would accounts for a factor 2, but a factor 4 leads to better agreement in simulations related to the shock tube, as reaction 2 is an influential reaction as shown further in the text. The ratio between the A-factors used for reactions 1 and 2 for xylene is then close to that used toluene. The addition of H-atoms to xylene molecules leads to the formation of toluene and methyl radicals (4), the addition of O-atoms gives H atoms and methylcresoxyl radicals (5) and the addition of OH radicals produces H-atoms and methylbenzyl alcohol (6). As in the case of toluene [24], an important part of the reactions of xylene molecules involves the formation of the resonance stabilized methylbenzyl radicals by unimolecular (1) and bimolecular (3) initiations and by metatheses with H-abstraction (7-23). The rate constants are directly deduced from those of the similar reactions in the case of toluene taking into account the number of abstractable H-atoms; this assumption has been proved correct in the case of the abstraction by H-atoms by a comparison with the experimental value proposed by Hippler *et al.* [49] (A = $4 \times 10^{14}$ and Ea = 8.4 kcal/mol compared to A = $4.8 \times 10^{14}$ and Ea = 8.4 kcal/mol with our assumption). We have also considered the formation of methyl tolyl radicals by abstraction of phenylic H-atoms (24-28), with rate constants deriving from those of the similar reactions in the case of benzene [11].

The reactions common to the three isomers of methylbenzyl radicals are directly derived from those of benzyl radicals : decomposition to form acetylene (29), reaction with oxygen molecules (31) and termination reactions with other radicals (34-39). By analogy with benzyl radicals, the decomposition (29) would lead to the formation of $\bullet C_5H_4CH_3$ radicals, but Emdee *et al.* [26] proposed a fast rearrangement of this species to give •H atoms and benzene. As we consider here only a high temperature mechanism to model results above 1100 K, we have neglected the formation of methylperoxybenzyl radicals. The reaction of methylbenzyl radicals with oxygen molecules lead to tolylmethoxyl radicals and O-atoms. Methylbenzyl radicals can combine with OH, $CH_3$, $HO_2$ and themselves, to give methylbenzylalcohol, ethyltoluene, methylbenzylhydroperoxide and bi-methylbenzyl, respectively. The reactions with •O• atoms, which are important at high temperature, lead to tolyl radicals and formaldehyde molecules (32) or to •H atoms and tolualdehyde molecules (35). Ortho- and para-methylbenzyl radicals can also react by decomposition (30) or with oxygen molecules (32) to form xylylenes. The rate constants of the decomposition by breaking of a C-H bond are those measured by Fernandes *et al.* [43] and those for the reactions with oxygen is based on the value proposed by Emdee *et al.* [26] for para-xylene. The activation energy (13.7 kcal/mol) proposed by Emdee *et al.* is consistent with the enthalpy of reaction which is about 10 kcal/mol at 1000 K. The formations of dihydronaphthalene and methylindene proposed by Marinov *et al.* [44] by addition to acetylene molecules of ortho-methylbenzyl radicals and meta- or para-methylbenzyl radicals, respectively, have also been taken into account (33).

The reactions of methyl tolyl radicals (40-46) have been adapted from those proposed for phenyl radicals [11] and includes reactions with oxygen molecules to give •O• atoms and methyl cresoxyl radicals and termination steps with other radicals.

Tolylmethoxyl radicals can react with oxygen to give tolualdehyde and •$HO_2$ radicals (49), by termination with H-atoms to give methyl benzyl alcohol (50) or by beta-scissions. Two beta-scissions are possible, the first involving the breaking of a C-H bond (47) to give tolualdehyde and H-atoms and the second involving the breaking of a C-C bond (48) to form formaldehyde and methyl phenyl radicals. As in the case of toluene, we have used the same activation energy for both channels.

The reaction of methyl cresoxyl radicals are the same as for cresoxyl radicals, i.e. CO elimination (51-52) and combination with H-atoms yielding methylcresol (53).

<u>Secondary mechanism</u>

Apart from the additions of H-atoms, which lead to •CHO radicals and toluene (57) and to benzaldehyde and methyl radicals (58), the reactions of tolualdehyde molecules involve the formation by initiations and by H-abstraction of two resonance stabilized radicals: methylbenzoyl radicals (formed by reactions 54, 55, 59-74), which decompose to methyl phenyl radicals and carbon monoxide (91), and methoxybenzyl radicals (formed by reaction 56, 74-90), which react according to the main steps of benzyl radicals (reactions 92-98) and lead to the formation of phthaladehyde or benzaldehyde.

Methyl benzyl hydroperoxyde molecules can easily decompose by rupture of the O-OH bond (99) with the rate constant proposed for the decomposition of benzyl hydroperoxyde molecules [24].

Methyl cresol molecules can react by addition of H-atoms, to give methyl radicals and cresols (102), or by bimolecular initiations (100 and 101) and by H-abstraction (103-118 and 119-134) to lead to the formation of two resonance stabilized radicals, methylcresoxyl and methylhydroxybenzyl radicals, respectively. The reactions deriving from methylhydroxybenzyl radicals (135-147) are deduced from the major steps deriving from benzyl radicals.

Apart from the additions of H-atoms, which lead to $CH_2OH$ radicals and toluene (149) or to methyl radicals and benzyl alcohol (150), the reactions of methylbenzylalcohol molecules involve the formation of the resonance stabilized tolylmethylhydroxyl radicals by bimolecular initiation (148) and by H-abstraction (151-166). The formation of the other possible radicals with a higher energy of formation has been neglected. Tolylmethylhydroxyl radicals can decompose to hydrogen atoms and tolualdehyde (167).

In the case of ethyltoluene molecules, the additions of H-atoms lead to $C_2H_5$ radicals and toluene (170) or to ethylbenzene and methyl radicals (171). Unimolecular (168), bimolecular initiation (169) and H-abstraction (172-187) involve the formation of the resonance stabilized tolylethyl radicals; the formation of the other possible radicals has been neglected, because they are not resonance stabilized. Tolylethyl radicals lead to the formation of methylstyrene by decompositions involving the breaking of a C-H bond (188) or by reaction with oxygen molecules (189) or combine with $HO_2$ radicals (190) to give hydroperoxide molecules, which quickly decompose.

In the case of bimethylbenzyl molecules, we have only considered the formation of the resonance stabilized ditolylethyl radicals by unimolecular (191) and bimolecular (192) initiation and by H-abstraction (193-208). Ditolylethyl radicals radicals can decompose to give hydrogen atoms and dimethylstilbene (209).

The reactions of xylylenes have not been much investigated previously. We consider that they can react with O-atoms to give methoxybenzyl radicals (212). Roth *et al.* [45] have studied the isomerization of o-xylylene to give benzocyclobutene (210) and proposed a rate constant that we have used. Benzocyclobutene leads to styrene (211) with a rate constant proposed by Tsang *et al.* [46].

Dihydronaphthalene leads to naphthalene through H-abstraction by H and O atoms and OH radicals (213-215) and beta-scission involving the breaking of a C-H bond (216).

## COMPARISON BETWEEN COMPUTED AND EXPERIMENTAL RESULTS

Simulations were performed using the CHEMKIN II software [32]. We have tried to reproduce our experimental data obtained in a shock tube, but to extend the validity of the proposed mechanism, we have also attempted to model results of the literature obtained in a flow tube [25, 26].

### Shock tubes

Computed results for our shock tube appear in figures 1 to 4. It is worth noting that the experimental OH emission at 306 nm is related to the electronically exited OH* population and is not directly proportional to OH radical concentration. Nevertheless, previous work [33] has shown a correct agreement between the shapes of the profiles of experimental emission and calculated OH concentration during the rise of the signal, which is the important part of the curve for the determination of ignition delays. Both experimentally and theoretically, the ignition delay is determined at 10% of the maximum OH peak. Simulations reproduce the systematic trends of the measurements, including variations in ignition delay times with temperature, equivalence ratio and concentration of hydrocarbon and that for the three isomers of xylenes. Simulations shown in fig. 4 reproduce well the fact that these three compounds have a very similar reactivity under these conditions.

### Flow reactor

Emdee *et al.* [25, 26] have studied the oxidation of the three isomers of xylene in a flow reactor at ~1150 K, at atmospheric pressure, with nitrogen as bath gas, for an initial concentration of xylene from 1400 to 1700 ppm and for equivalence ratios from 0.47 to 1.7.

Figures 5 to 7 display comparisons between the experimental and computed mole fraction of reactants and main products for the three xylenes for stochiometric mixtures. These figures show that a globally correct agreement can be observed. The consumptions of hydrocarbon and oxygen (fig. 5a, 6a, 7a) are correctly reproduced, even if that of meta-xylene is slightly underestimated. The formation of major products, tolualdehyde (fig. 5a, 6a, 7a), ethyl-toluene (fig. 5b, 6b, 7b), phthaldehyde (fig. 7b), methyl styrene (fig. 6b), toluene (fig. 5b, 6b, 7b), benzaldehyde (fig. 7b), benzene (fig. 5b, 6b), carbon monoxide (fig. 5a, 6a, 7a), methane (fig. 6b) and acetylene (fig. 7b) is also globally well captured by the simulations. In the case of ortho-xylene, the formation of styrene and naphthalene (fig. 5b), which is specific to this compound, is correctly simulated.

**FIGURES 5-7**

Figure 8 displays a comparison between the conversion of the three xylenes and shows that the proposed models correctly reproduce the difference of reactivity between these three compounds. Ortho-xylene is much more reactive than meta- and para-xylene, while para-xylene is slightly more reactive than meta-xylene. A comparison with the conversion of toluene computed under the same conditions as para-xylene shows that the monosubstituted compound is much less reactive for residence time below 40 ms and has a reactivity close to that of meta-xylene for larger residence times.

**FIGURE 8**

## ANALYSIS OF THE MECHANISM AND DISCUSSION

The flow and sensitivity analyses presented hereafter have been performed with the three mechanisms described previously. The accuracy of conclusions derived from them can only reflect the accuracy of the proposed reaction channels and of the used thermodynamic and kinetic data.

Figures 9 and 10 present sensitivity and flow analyses, respectively, computed in a flow reactor at 1155 K. Sensitivity analyses have been performed to know the effect of a variation of a given rate constant on the mole fraction of carbon monoxide. A positive coefficient indicates that the related reaction has a promoting effect. Sensitivity analyses are displayed for the three xylenes, but this flow analysis is only shown for ortho-xylene, as this isomer involves the highest numbers of reaction pathways including all those which are common to the three species. These analyses have been performed at the same residence time corresponding to a conversion of 36 % for ortho-xylene, 9.5% for meta-xylene and 18 % for para-xylene.

**FIGURES 9-10**

Figures 11 and 12 present sensitivity and flow analyses, respectively, computed in a shock tube. The analysis displayed in fig. 11 is for meta-xylene and shows the influence of the most sensitive reactions consuming xylene. Figure 11 presents also a comparison with toluene for a mixture with a close composition ($\Phi$ =1, 0.5% toluene compared to 0.625 % of xylene) and shows that this monosubstituted aromatic species is much less reactive than xylenes. The flux analysis displayed in fig. 12 has been computed for para-xylene at 1720 K.

**FIGURES 11-12**

## Main reaction pathways of xylene molecules

*Low temperature*

At 1155 K, in flow reactor conditions, whatever the isomers, xylene molecules mainly react according to four types of reactions in a very similar way compared to toluene:

- ♦ Ipso-additions of H-atoms to give methyl radicals and toluene (reaction 4 in Table V, 12 % of the consumption for ortho-xylene under the conditions of fig. 10),
- ♦ Abstractions of phenylic H-atoms mainly by OH radicals to give methyltolyl radicals (reactions 24, 13 % of the consumption for ortho-xylene), which mostly react with oxygen molecules to form methylcresoxy radicals,
- ♦ Ipso-additions of O-atoms to give H-atoms and methylcresoxy radicals (reaction 5, 13 % of the consumption for ortho-xylene), which mostly decompose to give carbon monoxide and benzene and methyl radicals or toluene and H-atoms,

- Abstractions of benzylic H-atoms mainly by H atoms and OH radicals to give methylbenzyl radicals (reactions 7 and 9, 60 % of the consumption for ortho-xylene).

Figure 9 shows that, at 1155 K, the ipso-additions of H-atoms have a slight inhibiting effect as they consume H-atoms to form the less reactive methyl radicals. The sensitivity on the formation of carbon monoxide of the ipso-additions of O-atoms (reaction 5) is probably enhanced by the fact that methylcresoxy radicals lead mainly to carbon monoxide through reactions 51 and 52. The abstractions of benzylic H-atoms reactions have a strong inhibiting effect as they lead to resonance stabilized methylbenzyl radicals, which react mainly by termination steps. To the contrary, the abstractions of phenylic H-atoms, which directly compete with the previous ones, have a strong promoting effect.

Initiation reactions (1, 2 or 3) are also present in the sensitivity analysis shown in fig. 9, even if their fluxes of consumption of xylene are negligible, apart from reaction (2) which accounts for 8% of the consumption of meta-xylene and strongly accelerates the oxidation of this species. Reaction (1) exhibits an important negative coefficient because the reverse termination step is dominant.

*High temperature*

At 1720 K, whatever the isomers, the flux analysis gives a much different picture: xylene molecules mainly react according to only three types of reactions :

- Ipso-additions of H-atoms to give methyl radicals and toluene (reaction 4, 6 % of the consumption for para-xylene under the conditions of fig. 12),
- Abstractions of benzylic H-atoms mainly by H atoms and OH radicals to give methylbenzyl radicals (reactions 7 and 9, 16 % of the consumption for para-xylene),
- Unimolecular initiation to give methyl radicals and tolyl radicals (reaction 2, 75 % of the consumption for para-xylene), which mainly react with oxygen molecules to give ultimately carbon monoxide and benzene.

Figure 11 displayed simulations obtained by dividing by 10 the rate constant of the unimolecular initiation (2) showing the strong promoting effect of this reaction at 1720 K. This figure displays also the effect of a division by 10 of the rate constant of the abstraction of benzylic H-atoms by H atoms (7) and OH radicals (9) enlightening the strong inhibiting effect of these reactions. The fact that the rate constant of the unimolecular initiation (1) is larger in the case of xylenes compared to toluene (a factor 6 at 1700K) mostly explains the lower reactivity of the monosubstituted aromatic compounds shown in fig. 11. While a study of the fast flow pyrolysis of p-xylene between 1200 and 1380 K [50] showed some importance of the H-abstraction between tolyl radicals and xylene, the flow rate of this reaction was found negligible under the conditions of fig. 10 and 12.

**Reactions deriving from the formation of methylbenzyl radicals and explanation of the difference of reactivity between the three isomers**

*Low temperature*

At 1155 K, in flow reactor conditions, as benzyl radicals, the three isomers of methylbenzyl radicals react mainly by termination steps with H-atoms (reaction –1), O-atoms (reaction 34, 35) and OH (reaction 36), methyl (reaction 38), and $HO_2$ (reaction 37) radicals. The termination reactions with H-atoms, OH and methyl radicals form, respectively, xylene, methyl benzylalcohol and ethyltoluene, which leads to methylstyrene, and have a slight inhibiting effect. A small fraction of the reaction with O-atoms leads to tolyl radicals (reaction 34), which react with oxygen molecules through a branching step to give methylphenoxy radicals and O-atoms. That explains the promoting effect of this reaction and the inhibiting one of the directly competing channel producing tolualdehyde and H-atoms (reaction 35). The termination steps with $HO_2$• radicals have a strong promoting effect as they form a hydroperoxide which almost immediately decomposes to give OH and tolylmethoxyl radicals (reaction 99). These last radicals in turn decompose to produce formaldehyde and tolyl radicals (reaction 48) or H-atoms and tolualdehyde (reaction 47). Tolualdehyde can react by abstraction by OH radicals of either a benzylic H-atom (reaction 77) to lead to benzaldehyde and formaldehyde or to phthaldehyde or a aldehydic H-atom (reaction 61) to involve the formation of benzene and carbon monoxide.

In the case of meta-methylbenzyl radicals, these termination steps are the only possible reactions and that explains the very low reactivity of meta-xylene, especially as the consumption of H-atoms by the termination step with methylbenzyl radicals has a higher flow rate than that by branching step with oxygen molecules. In the case of ortho- and para-methylbenzyl radicals, the formation of xylylene by reaction with oxygen molecules (reaction 32) is possible and has an important flux (20 % of the consumption of xylene). This reaction increases the concentration of $HO_2$ radicals, which is more than 10 times larger in the case of ortho-and para-xylene than for the meta isomer in the same conditions. This step favors the formation of methylbenzylhydroperoxide and promotes the global reactivity of ortho-and para-xylene as shown in fig. 8.

In summary, the reactions of meta-xylene are very similar to that of toluene, which explains why both compounds have a close reactivity for residence times above 40 ms in a flow reactor, as shown in fig. 8. Nevertheless, as the reversible recombinations of benzyl radicals with themselves ($\Delta H_r$(298K) = -64.1 kcal/mol) are thermodynamically favoured compared to that of methylbenzyl radicals (reaction 39) ($\Delta H_r$(298K) = -49.2 kcal/mol), their fluxes are larger for toluene (13.5% of the consumption benzyl radicals under the same conditions as in fig. 10) than for xylenes (0.04% of the consumption methylbenzyl radicals for meta-xylene). That explains why the conversion of toluene is much lower for low residence times, before an equilibrium bibenzyl concentration is reached.

Ortho-xylylene is consumed for around 50 % by addition of H-atoms to give back methylbenzyl radicals (reaction –30) and for 50 % by isomerisation to give benzocyclobutene (reaction 210) and then styrene (reaction 211); while para-xylylene is 80% consumed by addition of H-atoms to give back methylbenzyl radicals and only 20 % by reaction with O-atoms to give H-atoms and methoxybenzyl radicals (reaction 212). The consumption of H-atoms to give back the resonance stabilized methylbenzyl radicals decreases the global reactivity and the absence of possible concurrent isomerisation in the case of para-xylylene explains then why it is much less reactive than ortho-xylene as shown in fig. 8.

*High temperature*

At 1720 K, methylbenzyl radicals react mainly by termination steps with H-atoms (reaction –1) and with O-atoms to give tolualdehyde and H-atoms (reaction 35). The formation of para-xylylene under the conditions of figure 12 has a very small flux (2% of the consumption of xylene), which explains why there is no difference of reactivity in shock tube experiments, taking into account that the rate constants of the very sensitive initiation reaction (reaction 1) and abstractions of benzylic H-atoms (reactions 7 and 9) are the same whatever the isomer.

## CONCLUSION

This paper presents new experimental results for the autoignition of the three isomers of xylene in a shock tube between 1300 to 1820 K, showing a similar reactivity for the three compounds. It also proposes a set of detailed mechanisms able to reproduce these data, but also previously published experiments in a flow reactor at 1155 K [25, 26] exhibiting important differences of reactivity between the three species. To our knowledge, it is the first attempt of a kinetic detailed modeling of the three isomers of this bisubstituted aromatic species.

The reactions of importance in these mechanisms have been determined by using flux and sensitivity analyses. That has allowed us to explain the differences of reactivity obtained between the three isomers of xylenes: the possible ways of formation and consumption of xylylenes are responsible of the important differences of reactivity observed in a flow reactor at 1155 K, while the dominant importance of initiation reactions with the same rate constant encounters the fact that no difference is obtained in a shock tube between 1300 and 1800 K.

It is worth noting that these mechanisms are mostly based on reactions and rate constants guessed from analogies with toluene and benzene and should certainly be revisited when more experimental data are available. Future work could include the extension of the validity these mechanisms towards lower temperatures, i.e. below 1000 K, by considering the formation of peroxy radicals, in order to model the results obtained in a rapid compression machine [27, 28].

**ACKNOWLEDGEMENT**

This work has been supported by the European Commission by the contract EVG1-CT-2002-00072 SAFEKINEX.

## FIGURE CAPTIONS

Figure 1: **Auto-ignition delay times of ortho-xylene in a shock tube.** Semi-log plot of experimental (symbols) and computed (lines) ignition delay times as a function of temperature behind reflected shock wave (a) for different equivalence ratios at 0.375 % of xylene and (b) for different hydrocarbon concentrations at $\Phi = 1$.

Figure 2: **Auto-ignition delay times of meta-xylene in a shock tube.** Semi-log plot of experimental (symbols) and computed (lines) ignition delay times as a function of temperature behind reflected shock wave (a) for different equivalence ratios at 0.375 % of xylene and (b) for different hydrocarbon concentrations at $\Phi = 1$.

Figure 3: **Auto-ignition delay times of para-xylene in a shock tube.** Semi-log plot of experimental (symbols) and computed (lines) ignition delay times as a function of temperature behind reflected shock wave (a) for different equivalence ratios at 0.375 % of xylene and (b) for different hydrocarbon concentrations at $\Phi = 1$.

Figure 4: **Comparisons between the ignition delay times of the three xylenes.** Semi-log plot of experimental (symbols) and computed (lines) delay times as a function of temperature behind reflected shock wave at 0.375 % of xylene and $\Phi = 1$.

Figure 5: **Oxidation of o-xylene in a flow reactor at 1150 K, P= 1atm and $\Phi = 1.1$, with an initial concentration of hydrocarbon of 1700 ppm [26].** Comparison between experimental (symbols) and computed (lines) species mole fractions versus residence time with simulated time -10 ms shifted.

Figure 6: **Oxidation of m-xylene in a flow reactor at 1155 K, P= 1atm and $\Phi = 1.0$, with an initial concentration of hydrocarbon of 1580 ppm [26].** Comparison between experimental (symbols) and computed (lines) species mole fractions versus residence time with simulated time -20 ms shifted.

Figure 7: **Oxidation of p-xylene in a flow reactor at 1155 K, P= 1atm and Φ = 0.89, with an initial concentration of hydrocarbon of 1420 ppm [26].** Comparison between experimental (symbols) and computed (lines) species mole fractions versus residence time.

Figure 8: **Comparison between the conversions of the three xylenes and toluene in a flow reactor at 1155 K, P= 1atm and Φ close to 1 [26].**

Experimental (symbols) and computed (lines) conversions (only simulations for toluene) versus residence time under the same conditions as in fig. 5 to 7. Calculations have been performed for toluene under the same conditions as for para-xylene. The same time shifts as in fig. 5 and 6 have been used.

Figure 9: **Sensitivity analyses related to the mole fraction of CO in a flow reactor** at 1155 K, P= 1atm, Φ = 1.1 and a residence time of 0.02s. (for the clarity of the figure, only reactions of Table V, with an absolute value of the normalized sensitivity coefficient above 0.003 are shown).

Figure 10: **Flow analysis related to the oxidation of ortho-xylene in a flow reactor at 1155 K,** P= 1atm, Φ = 1.1 and a residence time of 0.02s. Products written in italics have not been experimentally observed [25].

Figure 11: **Sensitivity analyses in a shock tube for meta-xylene and comparison with toluene**. Semi-log plot of experimental (symbols) and computed (lines) ignition delay times as a function of temperature behind the reflected shock wave at Φ = 1. Mechanisms 2 and 3 have been obtained by dividing by 10 the rate constant of the initiation step (2) and that of the abstraction of benzylic H-atoms by H atoms (7) and OH radicals (9), respectively.

Figure 12: **Flow analysis related to the oxidation of para-xylene in a shock tube at 1720 K,** P=7.8 atm, $\Phi$ = 1 and a residence time of 9.5 $\mu$s corresponding to the autoignition start.

**TABLE I : Mixture compositions, shock conditions and ignition delay times for ortho-xylene**

| Composition (mole Percent) | | $P_1$ | P | V | T | τ | Composition (mole Percent) | | $P_1$ | P | V | T | τ |
|---|---|---|---|---|---|---|---|---|---|---|---|---|---|
| $C_8H_{10}$ | $O_2$ | (atm) | (atm) | (m/s) | (K) | (μs) | $C_8H_{10}$ | $O_2$ | (m/s) | (m/s) | (atm) | (K) | (s) |
| **0.375** | 3.9375 | 0.257 | 7.75 | 801 | 1446 | 361 | **0.375** | 1.96875 | 0.213 | 7.00 | 826 | 1545 | 159.2 |
| | | 0.268 | 8.12 | 802 | 1448 | 249 | | | 0.211 | 7.10 | 833 | 1566 | 84.5 |
| **Φ = 1** | | 0.233 | 7.42 | 817 | 1495 | 120 | **Φ = 2** | | 0.228 | 7.76 | 836 | 1576 | 82 |
| | | 0.245 | 8.00 | 824 | 1519 | 136 | | | 0.208 | 7.14 | 840 | 1589 | 71.3 |
| | | 0.226 | 7.58 | 831 | 1542 | 95 | | | 0.220 | 7.72 | 847 | 1613 | 70.5 |
| | | 0.229 | 7.76 | 835 | 1554 | 69.5 | | | 0.224 | 8.27 | 862 | 1666 | 32 |
| | | 0.221 | 8.11 | 859 | 1634 | 20.7 | | | 0.193 | 7.23 | 866 | 1678 | 25 |
| | | 0.216 | 8.13 | 457 | 1662 | 6.8 | | | 0.200 | 7.81 | 880 | 1727 | 11.3 |
| | | 0.204 | 7.98 | 878 | 1702 | 4.8 | | | 0.184 | 7.68 | 901 | 1803 | 7.3 |
| **0.375** | 7.875 | 0.263 | 7.22 | 774 | 1329 | 617 | **0.625** | 6.5625 | 0.266 | 8.07 | 794 | 1380 | 367 |
| | | 0.274 | 8.17 | 796 | 1397 | 116.7 | | | 0.257 | 7.99 | 802 | 1401 | 181.7 |
| **Φ = 0.5** | | 0.2341 | 7.26 | 807 | 1430 | 76.5 | **Φ = 1** | | 0.243 | 7.68 | 805 | 1410 | 175.8 |
| | | 0.245 | 7.77 | 814 | 1451 | 92.5 | | | 0.229 | 7.23 | 806 | 1412 | 170 |
| | | 0.230 | 7.35 | 815 | 1455 | 99.5 | | | 0.228 | 7.22 | 807 | 1416 | 189 |
| | | 0.220 | 7.03 | 817 | 1457 | 62 | | | 0.232 | 7.37 | 808 | 1419 | 130.7 |
| | | 0.223 | 7.33 | 823 | 1479 | 63.5 | | | 0.250 | 8.29 | 820 | 1455 | 179.3 |
| | | 0.211 | 7.8 | 859 | 1597 | 6 | | | 0.210 | 7.02 | 821 | 1459 | 67.5 |
| | | 0.217 | 8.26 | 868 | 1623 | 5.8 | | | 0.217 | 7.43 | 829 | 1483 | 84.7 |
| | | | | | | | | | 0.200 | 7.29 | 847 | 1541 | 17.5 |
| | | | | | | | | | 0.205 | 8.2 | 875 | 1631 | 9 |

*Note*: $P_1$ is the pressure of the mixture before the shock. V the speed of the incident wave. P and T are pressure and temperature behind the reflected shock wave. τ is the ignition delay time.

**TABLE II : Mixture compositions, shock conditions and ignition delay times for meta-xylene**

| Composition (mole Percent) | | $P_1$ | P | V | T | τ | Composition (mole Percent) | | $P_1$ | P | V | T | τ |
|---|---|---|---|---|---|---|---|---|---|---|---|---|---|
| $C_8H_{10}$ | $O_2$ | (atm) | (atm) | (m/s) | (K) | (μs) | $C_8H_{10}$ | $O_2$ | (m/s) | (m/s) | (atm) | (K) | (s) |
| **0.375** | 3.9375 | 0.246 | 7.26 | 795 | 1426 | 528 | **0.375** | 1.96875 | 0.250 | 7.33 | 794 | 1440 | 515 |
| | | 0.274 | 8.17 | 799 | 1437 | 469 | | | 0.258 | 7.71 | 804 | 1470 | 480 |
| **Φ = 1** | | 0.263 | 7.89 | 799 | 1439 | 299 | **Φ = 2** | | 0.211 | 0.257 | 8.04 | 1501 | 273 |
| | | 0.264 | 8.17 | 809 | 1469 | 288 | | | 0.250 | 8.08 | 822 | 1530 | 322 |
| | | 0.228 | 7.06 | 809 | 1471 | 205.8 | | | 0.243 | 7.91 | 823 | 1535 | 249 |
| | | 0.237 | 7.78 | 826 | 1524 | 71.5 | | | 0.2383 | 7.77 | 825 | 1540 | 180 |
| | | 0.250 | 8.32 | 829 | 1537 | 91.5 | | | 0.2173 | 7.11 | 826 | 1543 | 207 |
| | | 0.231 | 7.99 | 842 | 1575 | 61.5 | | | 0.237 | 8.21 | 843 | 1600 | 80.5 |
| | | 0.238 | 8.76 | 860 | 1636 | 15.5 | | | 0.217 | 7.43 | 848 | 1619 | 68 |
| | | 0.235 | 8.85 | 867 | 1663 | 17.5 | | | 0.189 | 6.65 | 849 | 1622 | 88 |
| | | 0.212 | 8.07 | 870 | 1675 | 14 | | | 0.243 | 8.82 | 856 | 1647 | 66.7 |
| | | 0.2032 | 7.76 | 872 | 1680 | 12.5 | | | 0.224 | 8.21 | 860 | 1658 | 48 |
| | | | | | | | | | 0.197 | 7.6 | 875 | 1712 | 24 |
| **0.375** | 7.875 | 0.270 | 8.06 | 797 | 1399 | 172.7 | **0.625** | 6.5625 | 0.270 | 7.74 | 780 | 1336 | 798 |
| | | 0.249 | 7.45 | 798 | 1402 | 279 | | | 0.230 | 6.82 | 787 | 1359 | 407 |
| **Φ = 0.5** | | 0.232 | 7.27 | 811 | 1442 | 136 | **Φ = 1** | | 0.235 | 7.35 | 802 | 1401 | 308 |
| | | 0.243 | 7.66 | 812 | 1444 | 104 | | | 0.250 | 7.82 | 804 | 1406 | 337 |
| | | 0.257 | 8.1 | 813 | 1446 | 90.8 | | | 0.233 | 7.45 | 810 | 1425 | 161.7 |
| | | 0.230 | 7.34 | 815 | 1453 | 86.7 | | | 0.213 | 7.04 | 819 | 1452 | 157.5 |
| | | 0.233 | 7.58 | 821 | 1472 | 58 | | | 0.243 | 8.06 | 820 | 1455 | 167.5 |
| | | 0.197 | 6.85 | 840 | 1533 | 15.5 | | | 0.263 | 8.76 | 821 | 1459 | 205.8 |
| | | 0.204 | 7.11 | 841 | 1538 | 22 | | | 0.193 | 6.9 | 840 | 1519 | 70.5 |
| | | 0.217 | 7.57 | 841 | 1538 | 22.7 | | | 0.210 | 7.46 | 842 | 1524 | 47.5 |
| | | 0.224 | 8.2 | 856 | 1586 | 22.5 | | | 0.217 | 7.82 | 844 | 1531 | 60.5 |
| | | 0.230 | 8.55 | 860 | 1600 | 10 | | | 0.199 | 7.22 | 847 | 1539 | 80 |
| | | 0.197 | 7.93 | 885 | 1681 | 3.5 | | | 0.199 | 7.87 | 873 | 1623 | 18.0 |
| | | | | | | | | | 0.204 | 8.1 | 874 | 1626 | 18.5 |
| | | | | | | | | | 0.210 | 8.40 | 875 | 1631 | 17 |
| | | | | | | | | | 0.204 | 8.58 | 892 | 1685 | 6.5 |

*Note*: $P_1$ is the pressure of the mixture before the shock. V the speed of the incident wave. P and T are pressure and temperature behind the reflected shock wave. τ is the ignition delay time.

**TABLE III : Mixture compositions, shock conditions and ignition delay times for para-xylene**

| Composition (mole Percent) | | $P_1$ | P | V | T | τ | Composition (mole Percent) | | $P_1$ | P | V | T | τ |
|---|---|---|---|---|---|---|---|---|---|---|---|---|---|
| $C_8H_{10}$ | $O_2$ | (atm) | (atm) | (m/s) | (K) | (μs) | $C_8H_{10}$ | $O_2$ | (m/s) | (m/s) | (atm) | (K) | (s) |
| **0.375** | 3.9375 | 0.254 | 7.93 | 811 | 1476 | 136 | **0.375** | 1.96875 | 0.224 | 7.03 | 814 | 1503 | 258 |
| | | 0.210 | 6.90 | 825 | 1522 | 94 | | | 0.200 | 6.39 | 818 | 1518 | 189 |
| **Φ = 1** | | 0.228 | 7.47 | 825 | 1522 | 100.7 | **Φ = 2** | | 0.203 | 6.57 | 823 | 1533 | 148 |
| | | 0.251 | 8.3 | 827 | 1529 | 93 | | | 0.237 | 7.9 | 831 | 1561 | 135.3 |
| | | 0.253 | 8.56 | 835 | 1554 | 80 | | | 0.213 | 7.15 | 833 | 1566 | 137.3 |
| | | 0.240 | 8.28 | 841 | 1572 | 64 | | | 0.226 | 7.62 | 835 | 1574 | 123.3 |
| | | 0.243 | 8.61 | 848 | 1597 | 38 | | | 0.230 | 7.95 | 841 | 1595 | 77.7 |
| | | 0.204 | 7.39 | 856 | 1622 | 21.5 | | | 0.218 | 7.61 | 845 | 1608 | 79.3 |
| | | 0.245 | 8.92 | 857 | 1628 | 15 | | | 0.221 | 7.98 | 855 | 1641 | 50 |
| | | 0.233 | 8.51 | 859 | 1634 | 18.5 | | | 0.195 | 7.08 | 857 | 1652 | 38.5 |
| | | 0.201 | 7.54 | 865 | 1657 | 12.3 | | | 0.205 | 7.8 | 871 | 1697 | 16 |
| | | 0.197 | 7.45 | 867 | 1663 | 10 | | | 0.197 | 7.65 | 877 | 1718 | 10 |
| | | 0.197 | 7.84 | 883 | 1720 | 8.3 | | | | | | | |
| | | | | | | | | | | | | | |
| **0.375** | 7.875 | 0.251 | 8.01 | 815 | 1453 | 102 | **0.625** | 6.5625 | 0.257 | 7.24 | 776 | 1324 | 540 |
| | | 0.225 | 7.36 | 823 | 1477 | 102 | | | 0.275 | 8.11 | 787 | 1359 | 343 |
| **Φ = 0.5** | | 0.217 | 7.2 | 826 | 1490 | 89.5 | **Φ = 1** | | 0.239 | 7.26 | 794 | 1381 | 164.2 |
| | | 0.238 | 8.1 | 835 | 1517 | 81.3 | | | 0.230 | 7.42 | 812 | 1431 | 287 |
| | | 0.201 | 7.02 | 841 | 1538 | 56 | | | 0.220 | 7.11 | 813 | 1435 | 112 |
| | | 0.224 | 7.8 | 841 | 1538 | 48.5 | | | 0.239 | 7.87 | 818 | 1450 | 162.7 |
| | | 0.217 | 7.89 | 853 | 1578 | 20.5 | | | 0.224 | 7.8 | 834 | 1500 | 62 |
| | | 0.237 | 8.72 | 857 | 1592 | 14.3 | | | 0.196 | 6.97 | 840 | 1519 | 59 |
| | | 0.210 | 7.88 | 863 | 1608 | 16 | | | 0.225 | 8.30 | 851 | 1554 | 21 |
| | | | | | | | | | 0.227 | 8.42 | 852 | 1557 | 23.5 |
| | | | | | | | | | 0.217 | 8.1 | 855 | 1564 | 27.5 |
| | | | | | | | | | 0.201 | 7.81 | 866 | 1601 | 11.3 |

*Note*: $P_1$ is the pressure of the mixture before the shock. V the speed of the incident wave. P and T are pressure and temperature behind the reflected shock wave. τ is the ignition delay time.

**TABLE IV: Names, formulae and heats of formation for aromatic species involved in Table V. Apart from xylenes, xylylenes and methylbenzyl radicals, only the compounds related to ortho-xylene are presented. The heats of formation have been calculated by software THERGAS [37] at 298 K in kcal.mol$^{-1}$.**

| Species | Structure | $\Delta H_f$ (298K) | Species | Structure | $\Delta H_f$ (298K) |
|---|---|---|---|---|---|
| o-xylene ($CH_3C_6H_4CH_3$) | 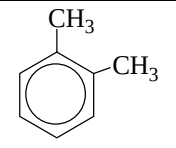 | 4.5 | Benzocyclobutene ($cCH_2\ CH_2C_6H_4$) | 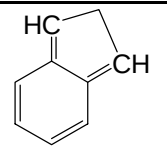 | 14.2 |
| m-xylene ($CH_3C_6H_4CH_3$) | 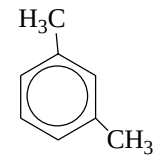 | 4.1 | o-methylbenzyl ($CH_2C_6H_4CH_3$) | 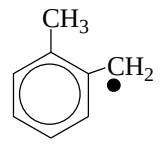 | 42.0 |
| p-xylene ($CH_3C_6H_4CH_3$) | 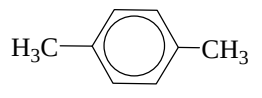 | 4.2 | m-methylbenzyl ($CH_2C_6H_4CH_3$) | 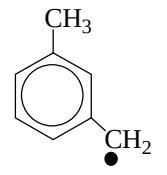 | 41.6 |
| Tolualdehyde ($CH_3C_6H_4CHO$) | 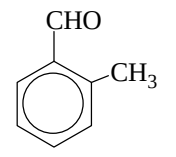 | -16.2 | p-methylbenzyl ($CH_2C_6H_4CH_3$) | 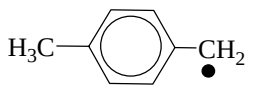 | 41.8 |
| Methyl benzyl hydropreroxide ($CH_3C_6H_4CH_2OOH$) | 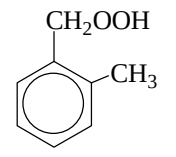 | -15.6 | Methyl Cresoxyl ($OC_6H_3(CH_3)_2$) | 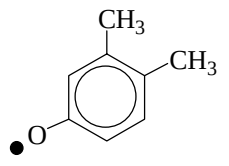 | -1.7 |
| Methylcresol ($HOC_6H_3(CH_3)_2$) | 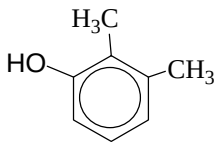 | -37.6 | Methyl tolyl ($CH_3C_6H_3CH_3$) | 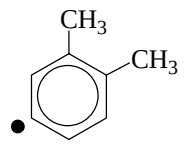 | 65.7 |
| Methylbenzyl alcohol ($CH_3C_6H_4CH_2OH$) | 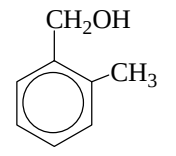 | -1.5 | Tolyl Methoxyl ($CH_3C_6H_4CH_2O$) | 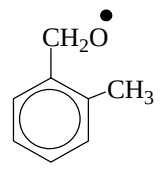 | 19.9 |
| Ethyl toluene ($CH_3C_6H_4CH_2CH_3$) | 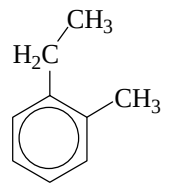 | -0.1 | Methyl benzoyl ($CH_3C_6H_4CO$) | 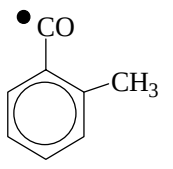 | 18.6 |
| Methylstyrene ($CH_3C_6H_5CH{=}CH_2$) | 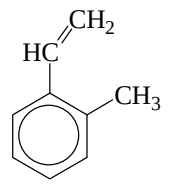 | 28.4 | Methoxy benzyl ($CH_2C_6H_4CHO$) | 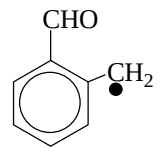 | 21.4 |
| Bimethylbenzyl ($C_{16}H_{18}$) | 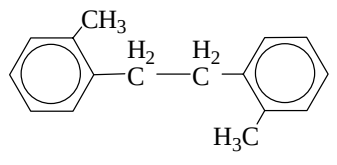 | 34.0 | Benzaldehyde Methoxyl ($CH_2OC_6H_4CHO$) | 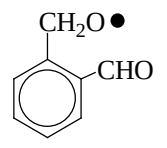 | 1.1 |
| Mesitylene ($C_6H_3(CH_3)_3$) | 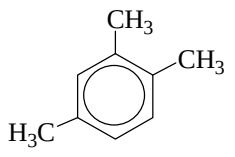 | -1.8 | Methyl hydroxy benzyl ($HOC_6H_3CH_3CH_2$) | 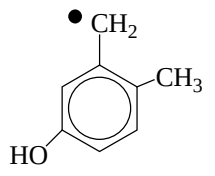 | -0.1 |
| Dihydronaphthalene ($C_{10}H_{10}$) | 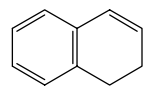 | 33.2 | Tolyl methyl hydroxyl ($CH_3C_6H_4CHOH$) | 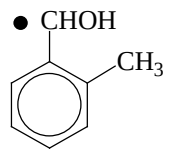 | 1.9 |
| Naphthalene ($C_{10}H_8$) | 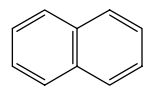 | 36.0 | Hydroxy tolyl methoxyl ($HOC_6H_3CH_3CH_2O$) | 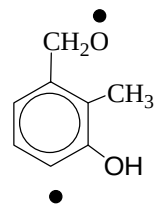 | -22.9 |
| o-xylylene ($CH_2C_6H_4CH_2$) | 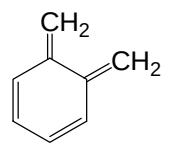 | 53.0[a] | Methyl hydroxy benzoyl ($HOC_6H_4CH_3CO$) | 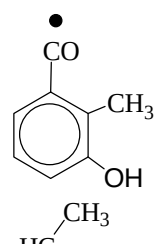 | -24.2 |
| p-xylylene ($CH_2C_6H_4CH_2$) | 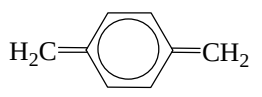 | 50.1[a] | Tolyl ethyl ($CH_3C_6H_4CHCH_3$) | 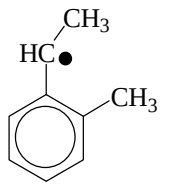 | 33.3 |
| Phthaldehyde ($CHOC_6H_4CHO$) | 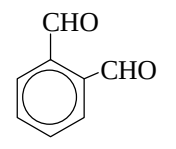 | -35.1 | Ditolyl ethyl ($C_{16}H_{17}$) | 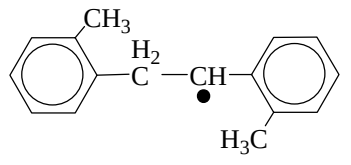 | 67.3 |
| Dimethyl stilbene ($CH_3C_6H_4CH{=}CHC_6H_4CH_3$) | 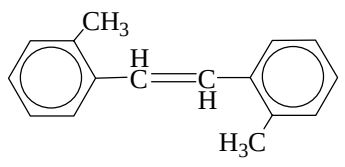 | 44.2 | Naphthyl ($C_{10}H_9$) | 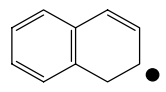 | 60.1 |
| Hydroxy tolualdehyde ($CH_3C_6H_4(OH)CHO$) | 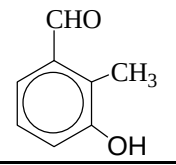 | -59.0 | | | |

[a] : Value taken from Pollack *et al.* [41].

**TABLE V : Primary and secondary mechanism for the oxidation of xylenes**

The rate constants are given at 1 atm (k=A $T^n$ exp(-$E_a$/RT)) in cc, mol, s, kcal units. Reference numbers are given in brackets when they appear for the first time. This mechanism should be used together with the mechanism for the oxidation of benzene [11] and toluene [24]. When the type of isomer is not specified, the reaction considered in the mechanism for each of the three compounds.

| Reactions | A | n | $E_a$ | References | Rxn No |
|---|---|---|---|---|---|
| **PRIMARY MECHANISM** | | | | | |
| **Reactions of xylene molecules** | | | | | |
| *Umimolecular inititiation* | | | | | |
| o-xylene=o-methylbenzyl+H | $5.0\times10^{15}$ | 0.0 | 88.3 | DACOSTA00[42][a] | (1) |
| m-xylene=m-methylbenzyl+H | $5.0\times10^{15}$ | 0.0 | 89.7 | DACOSTA00[42][a] | (1) |
| p-xylene=p-methylbenzyl+H | $5.0\times10^{15}$ | 0.0 | 89.0 | DACOSTA00[42][a] | (1) |
| xylene=C6H4CH3+CH3 | $4.0\times10^{17}$ | 0.0 | 97.0 | estimated[b] | (2) |
| *Bimolecular initiation* | | | | | |
| xylene+O2=methylbenzyl+HO2 | $4.2\times10^{12}$ | 0.0 | 38.6 | estimated[c] | (3) |
| *additions* | | | | | |
| xylene+H=toluene+CH3 | $11.6\times10^{13}$ | 0.0 | 8.1 | estimated[d] | (4) |
| xylene+O=OC6H3(CH3)2+H | $1.7\times10^{13}$ | 0.0 | 3.6 | estimated[e] | (5) |
| xylene+OH=HOC6H3(CH3)2+H | $1.3\times10^{13}$ | 0.0 | 10.6 | estimated[e] | (6) |
| *Metatheses* | | | | | |
| *Metatheses with abstraction of a methylbenzylic H-atom* | | | | | |
| xylene+R=methylbenzyl+RH | | | | estimated[d] | (7-23) |
| R is H, O, OH, HO2, CH3, C2H3, C3H5, C3H3, I-C4H5, nC4H5, cyclopentadienyl, phenyl, phenoxyl, benzyl, cresoxyl, tolyl, methylcresoxyl) | | | | | |
| *Metatheses with abstraction of a phenylic H-atom* | | | | | |
| xylene+R=methyltolyl+RH | | | | estimated[f] | (24-28) |
| R is H, O, OH, HO2, CH3 | | | | | |
| **Reactions of methylbenzyl radicals** | | | | | |
| *Decompositions by beta-scission* | | | | | |
| methylbenzyl=>C6H6+H+C2H2 | $6.0\times10^{13}$ | 0.0 | 70.0 | estimated[e] | (29) |
| o-methylbenzyl=o-xylylene+H | $5.0\times10^{15}$ | 0.0 | 74.2 | FERNANDES02[43] | (30) |
| p-methylbenzyl=p-xylylene+H | $5.0\times10^{15}$ | 0.0 | 70.6 | FERNANDES02[43] | (30) |
| *Reactions with oxygen molecules* | | | | | |
| methylbenzyl+O2=CH3C6H4CH2O+O | $6.3\times10^{12}$ | 0.0 | 40.0 | estimated[e] | (31) |
| o-methylbenzyl+O2=>o-xylylene+HO2 | $3.6\times10^{13}$ | 0.0 | 16.7 | estimated[g] | (32) |
| p-methylbenzyl+O2=>p-xylylene+HO2 | $3.6\times10^{13}$ | 0.0 | 13.7 | EMDEE91[26] | (32) |
| *Addition reactions* | | | | | |
| o-methylbenzyl+C2H2=>C10H10+H | $3.2\times10^{11}$ | 0.0 | 7.0 | MARINOV96[44] | (33) |
| m-methylbenzyl+C2H2=>methylindene+H | $3.2\times10^{11}$ | 0.0 | 7.0 | MARINOV96 | (33) |
| p-methylbenzyl+C2H2=>methylindene+H | $3.2\times10^{11}$ | 0.0 | 7.0 | MARINOV96 | (33) |
| *Termination reactions* | | | | | |
| methylbenzyl+O=tolyl+HCHO | $3.5\times10^{13}$ | 0.0 | 0.0 | estimated[e] | (34) |
| methylbenzyl+O=CH3C6H4CHO+H | $1.0\times10^{14}$ | 0.0 | 0.0 | estimated[e] | (35) |
| methylbenzyl+OH=CH3C6H4CH2OH | $2.0\times10^{13}$ | 0.0 | 0.0 | estimated[e] | (36) |
| methylbenzyl+HO2=CH3C6H4CH2OOH | $5.0\times10^{12}$ | 0.0 | 0.0 | estimated[e] | (37) |
| etC6H4CH3=methylbenzyl+CH3 | $6.1\times10^{15}$ | 0.0 | 75.1 | estimated[e] | (38) |
| 2methylbenzyl=bimethylbenzyl | $2.5\times10^{11}$ | 0.4 | 0.0 | estimated[e] | (39) |
| **Reactions of methyltolyl radicals** | | | | | |
| *Reactions with oxygen molecule* | | | | | |
| methyltolyl+O2=OC6H4(CH3)2+O | $2.6\times10^{13}$ | 0.0 | 6.1 | estimated[e] | (40) |
| *Termination reactions* | | | | | |
| methyltolyl+H=xylene | $1.0\times10^{14}$ | 0.0 | 0.0 | estimated[e] | (41) |
| methyltolyl+H=methylbenzyl+H | $2.0\times10^{13}$ | 0.0 | 0.0 | estimated[d] | (42) |
| methyltolyl+O=OC6H3(CH3)2 | $1.0\times10^{14}$ | 0.0 | 0.0 | estimated[e] | (43) |
| methyltolyl+OH=HOC6H4(CH3)2 | $1.0\times10^{13}$ | 0.0 | 0.0 | estimated[e] | (44) |
| methyltolyl+CH3=mesitylene | $1.2\times10^{6}$ | 1.96 | -3.7 | estimated[e] | (45) |
| methyltolyl+HO2=OC6H3(CH3)2+OH | $5.0\times10^{12}$ | 0.0 | 0.0 | estimated[e] | (46) |
| **Reactions of tolylmethoxyl radicals** | | | | | |
| *Decomposition by beta-scission* | | | | | |
| CH3C6H4CH2O=H+CH3C6H4CHO | $2.0\times10^{13}$ | 0.0 | 27.5 | estimated[e] | (47) |
| CH3C6H4CH2O=C6H4CH3+HCHO | $2.0\times10^{13}$ | 0.0 | 27.5 | estimated[e] | (48) |
| *Reactions with oxygen* | | | | | |
| CH3C6H4CH2O+O2=HO2+CH3C6H4CHO | $6.0\times10^{10}$ | 0.0 | 1.6 | estimated[e] | (49) |
| *Termination reactions* | | | | | |
| CH3C6H4CH2O+H=CH3C6H4CH2OH | $1.0\times10^{14}$ | 0.0 | 0.0 | estimated[h] | (50) |

**Reactions of methylcresoxyl radicals**

| | | | | | |
|---|---|---|---|---|---|
| *CO elimination with rearrangement* | | | | | |
| OC6H3(CH3)2=H+toluene+CO | $3.0\times10^{11}$ | 0.0 | 43.8 | estimated[e] | (51) |
| OC6H3(CH3)2=CH3+benzene+CO | $3.0\times10^{11}$ | 0.0 | 43.8 | estimated[e] | (52) |
| *Termination reactions* | | | | | |
| OC6H3(CH3)2+H=HOC6H3(CH3)2 | $1.0\times10^{14}$ | 0.0 | 0.0 | estimated[e] | (53) |

## SECONDARY MECHANISM

**Reactions of tolualdehyde molecules and derived radicals**

| | | | | | |
|---|---|---|---|---|---|
| CH3C6H4CH0=CH3C6H4C0+H | $4.0\times10^{15}$ | 0.0 | 83.7 | estimated[e] | (54) |
| CH3C6H4CH0+O2=CH3C6H4C0+HO2 | $7.0\times10^{11}$ | 0.0 | 39.5 | estimated[c] | (55) |
| CH3C6H4CH0+O2=CH2C6H4CH0+HO2 | $2.1\times10^{12}$ | 0.0 | 41.0 | estimated[c] | (56) |
| CH3C6H4CH0+H=toluene+CHO | $5.8\times10^{13}$ | 0.0 | 8.1 | estimated[e] | (57) |
| CH3C6H4CH0+H=C6H5CHO+CH3 | $5.8\times10^{13}$ | 0.0 | 8.1 | estimated[i] | (58) |
| CH3C6H4CH0+R=CH3C6H4C0+RH | | | | estimated[e] | (59-74) |
| CH3C6H4CH0+R=CH2C6H4CH0+RH | | | | estimated[i] | (75-90) |
| R is H, O, OH, HO2, CH3, C2H5, C3H5, C3H3, i-C4H5, nC4H5, cyclopentadienyl, phenoxyl, benzyl, cresoxyl, methylbenzyl, OC6H4(CH3)2 | | | | | |
| CH3C6H4C0=C6H4CH3+CO | $4.0\times10^{14}$ | 0.0 | 29.5 | estimated[e] | (91) |
| CH2C6H4CH0+O2=CH2OC6H4CHO+O | $6.3\times10^{12}$ | 0.0 | 40.0 | estimated[i] | (92) |
| CH2C6H4CHO+H= CH3C6H4CHO | $1.0\times10^{14}$ | 0.0 | 0.0 | estimated[i] | (93) |
| CH2C6H4CHO+HO2=CH2OC6H4CHO+OH | $5.0\times10^{12}$ | 0.0 | 0.0 | estimated[i] | (94) |
| CH2C6H4CHO+CH3=C6H5CHO+C2H4 | $5.0\times10^{12}$ | 0.0 | 0.0 | estimated[i] | (95) |
| CH2OC6H4CHO=>H+C6H4(CHO)2 | $2.0\times10^{13}$ | 0.0 | 27.5 | estimated[i] | (96) |
| CH2OC6H4CHO=C6H5CHO+HCO | $2.0\times10^{13}$ | 0.0 | 27.5 | estimated[i] | (97) |
| CH2OC6H4CHO+O2=>OOH+C6H4(CHO)2 | $6.0\times10^{10}$ | 0.0 | 1.6 | estimated[i] | (98) |

**Reactions of methylbenzyl hydropreroxyde molecules**

| | | | | | |
|---|---|---|---|---|---|
| C6H5CH2OOH=C6H5CH2O+OH | $1.5\times10^{16}$ | 0.0 | 42.0 | estimated[e] | (99) |

**Reactions of methylcresol molecules and derived radicals**

| | | | | | |
|---|---|---|---|---|---|
| HOC6H3(CH3)2+O2=OC6H3(CH3)2+HO2 | $1.0\times10^{13}$ | 0.0 | 38.9 | estimated[c] | (100) |
| HOC6H3(CH3)2+O2=HOC6H3CH3CH2+HO2 | $2.1\times10^{12}$ | 0.0 | 39.7 | estimated[c] | (101) |
| HOC6H3(CH3)2+H=HOC6H4CH3+CH3 | $5.8\times10^{13}$ | 0.0 | 8.1 | estimated[e] | (102) |
| HOC6H3(CH3)2+R=OC6H3(CH3)2+RH | | | | estimated[e] | (103-118) |
| HOC6H3(CH3)2+R=HOC6H3CH3CH2+RH | | | | estimated[d] | (119-134) |
| R is H, O, OH, HO2, CH3, C2H5, C3H5, C3H3, i-C4H5, nC4H5, cyclopentadienyl, phenoxyl, benzyl, cresoxyl, methylbenzyl, OC6H4(CH3)2 | | | | | |
| HOC6H3CH3CH2+O2=HOC6H3CH3CH2O+O | $6.3\times10^{12}$ | 0.0 | 40.0 | estimated[i] | (135) |
| HOC6H3CH3CH2+H=HOC6H3(CH3)2 | $1.0\times10^{14}$ | 0.0 | 0.0 | estimated[i] | (136) |
| HOC6H3CH3CH2+HO2=HOC6H3CH3CH2O+OH | $5.0\times10^{12}$ | 0.0 | 0.0 | estimated[i] | (137) |
| HOC6H3CH3CH2+CH3=HOC6H4CH3+C2H4 | $5.0\times10^{12}$ | 0.0 | 0.0 | estimated[i] | (138) |
| HOC6H3CH3CH2O=H+HOC6H3CH3CHO | $2.0\times10^{13}$ | 0.0 | 27.5 | estimated[i] | (139) |
| HOC6H3CH3CH2O=OC6H4CH3+HCHO | $2.0\times10^{13}$ | 0.0 | 27.5 | estimated[i] | (140) |
| HOC6H3CH3CH2O+O2=HO2+HOC6H3CH3CHO | $6.0\times10^{10}$ | 0.0 | 1.6 | estimated[i] | (141) |
| HOC6H3CH3CHO+R=HOC6H3CH3CO+RH | $4.0\times10^{13}$ | 0.0 | 3.2 | estimated[i] | (142-146) |
| R is H, O, OH, HO2, CH3 | | | | | |
| HOC6H3CH3CO=OC6H4CH3+CO | $4.0\times10^{14}$ | 0.0 | 29.5 | estimated[d] | (147) |

**Reactions of methylbenzylalcohol molecules and derived radicals**

| | | | | | |
|---|---|---|---|---|---|
| CH3C6H4CH2OH+O2=HO2+CH3C6H4CHOH | $1.4\times10^{12}$ | 0.0 | 34.0 | estimated[c] | (148) |
| CH3C6H4CH2OH+H=toluene+CH2OH | $5.8\times10^{13}$ | 0.0 | 8.1 | estimated[e] | (149) |
| CH3C6H4CH2OH+H=CH3+C6H5CH2OH | $5.8\times10^{13}$ | 0.0 | 8.1 | estimated[i] | (150) |
| CH3C6H4CH2OH+R=RH+CH3C6H4CHOH | | | | estimated[e] | (151-166) |
| R is H, O, OH, HO2, CH3, C2H5, C3H5, C3H3, i-C4H5, nC4H5, cyclopentadienyl, phenoxyl, benzyl, cresoxyl, methylbenzyl, OC6H4(CH3)2 | | | | | |
| CH3C6H4CHOH=CH3C6H4CHO+H | 2.0x10$^{14}$ | 0.0 | 23.3 | estimated[e] | (167) |
| **Reactions of ethyltoluene molecules and derived radicals** | | | | | |
| etC6H4CH3=CH3CHC6H4CH3+H | 4.3x10$^{14}$ | 0.0 | 83.6 | estimated[e] | (168) |
| etC6H4CH3+O2=CH3CHC6H4CH3+HO2 | 1.4x10$^{12}$ | 0.0 | 34.0 | estimated[c] | (169) |
| etC6H4CH3+H=toluene+C2H5 | 5.8x10$^{13}$ | 0.0 | 8.1 | estimated[e] | (170) |
| etC6H4CH3+H=etC6H5+CH3 | 5.8x10$^{13}$ | 0.0 | 8.1 | estimated[e] | (171) |
| etC6H4CH3+R=CH3CHC6H4CH3+RH | | | | estimated[e] | (172-187) |
| R is H, O, OH, HO2, CH3, C2H5, C3H5, C3H3, i-C4H5, nC4H5, cyclopentadienyl, phenoxyl, benzyl, cresoxyl, methylbenzyl, OC6H4(CH3)2 | | | | | |
| CH3CHC6H4CH3=H+methylstyrene | 3.1x10$^{13}$ | 0.0 | 50.7 | estimated[e] | (188) |
| CH3CHC6H4CH3+O2=HO2+methylstyrene | 7.0x10$^{11}$ | 0.0 | 15.0 | estimated[e] | (189) |
| CH3CHC6H4CH3+HO2=OH+CH3C6H4CHO+CH3 | 5.0x10$^{12}$ | 0.0 | 0.0 | estimated[e] | (190) |
| **Reactions of bimethylbenzyl molecules and derived radicals** | | | | | |
| Bimethylbenzyl=C16H17+H | 1.0x10$^{14}$ | 0.0 | 83.0 | estimated[e] | (191) |
| bimethylbenzyl+O2=C16H17+HO2 | 2.8x10$^{12}$ | 0.0 | 35.0 | estimated[e] | (192) |
| bimethylbenzyl+R=C16H17+RH | | | | estimated[e] | (193-208) |
| R is H, O, OH, HO2, CH3, C2H5, C3H5, C3H3, i-C4H5, nC4H5, cyclopentadienyl, phenoxyl, benzyl, cresoxyl, methylbenzyl, OC6H4(CH3)2 | | | | | |
| C16H17=dimethylstilbene+H | 7.1x10$^{14}$ | 0.0 | 30.0 | estimated[e] | (209) |
| **Reactions of xylylene molecules and derived compounds (only for ortho- or para-xylene)** | | | | | |
| o-xylylene=benzocyclobutene | 2.1x10$^{12}$ | 0.0 | 26.8 | ROTH81[45] | (210) |
| benzocyclobutene=styrene | 1.2x10$^{15}$ | 0.0 | 74.3 | TSANG90[46] | (211) |
| o-xylylene+O=o-CH2C6H4CHO+H | 6.0x10$^{8}$ | 1.45 | 0.9 | estimated[j] | (212) |
| p-xylylene+O=o-CH2C6H4CHO+H | 6.0x10$^{8}$ | 1.45 | 0.9 | estimated[j] | (212) |
| **Reactions of dihydronaphthalene molecules and derived compounds (only for ortho-xylene)** | | | | | |
| C10H10+OH=C10H9+H2O | 5.0x10$^{6}$ | 2.0 | 0.0 | MARINOV96 | (213) |
| C10H10+H=C10H9+H2 | 2.0x10$^{5}$ | 2.5 | 2.5 | MARINOV96 | (214) |
| C10H10+O=C10H9+OH | 7.0x10$^{11}$ | 0.7 | 6.0 | MARINOV96 | (215) |
| naphthalene+H=C10H9 | 5.0x10$^{14}$ | 0.0 | 5.0 | MARINOV96 | (216) |

a : The A factor has been divided by 2.

b : Rate constant taken equal to that proposed for toluene [24] multiplied by 4.

c: The rate constant of this bimolecular initiation with oxygen molecule has been calculated as proposed by Ingham *et al* [47] : A is taken equal to n x $7x10^{11}$ $cm^3.mol^{-1}s^{-1}$, where n is the number of benzylic hydrogen atoms abstractable, and the activation energy to the reaction enthalpy.

d: Rate constant taken equal to that of the similar reaction in the mechanism of the oxidation of toluene [24] multiplied by 2 to take into account the number of abstractable H-atoms or of methyl groups.

e: Rate constant taken equal to that of the similar reaction in the mechanism of the oxidation of toluene [24].

f: Rate constant taken equal to that proposed for benzene [11] multiplied by 2.

g: Rate constant deduced from that of the similar reaction in the case of p-xylene and the difference in enthalpies of formation between the ortho and para isomers of xylylene [41].

h: Rate constant taken equal to that of the recombination of •H atoms with alkyl radicals as proposed by Allara *et al.* [48].

i: Rate constant taken equal to that of the similar reaction in the case of toluene or benzyl radicals and derived radicals [24].

j: Rate constant taken equal to that of the similar reaction in the case of 1,3-butadiene [33].

Figure 1

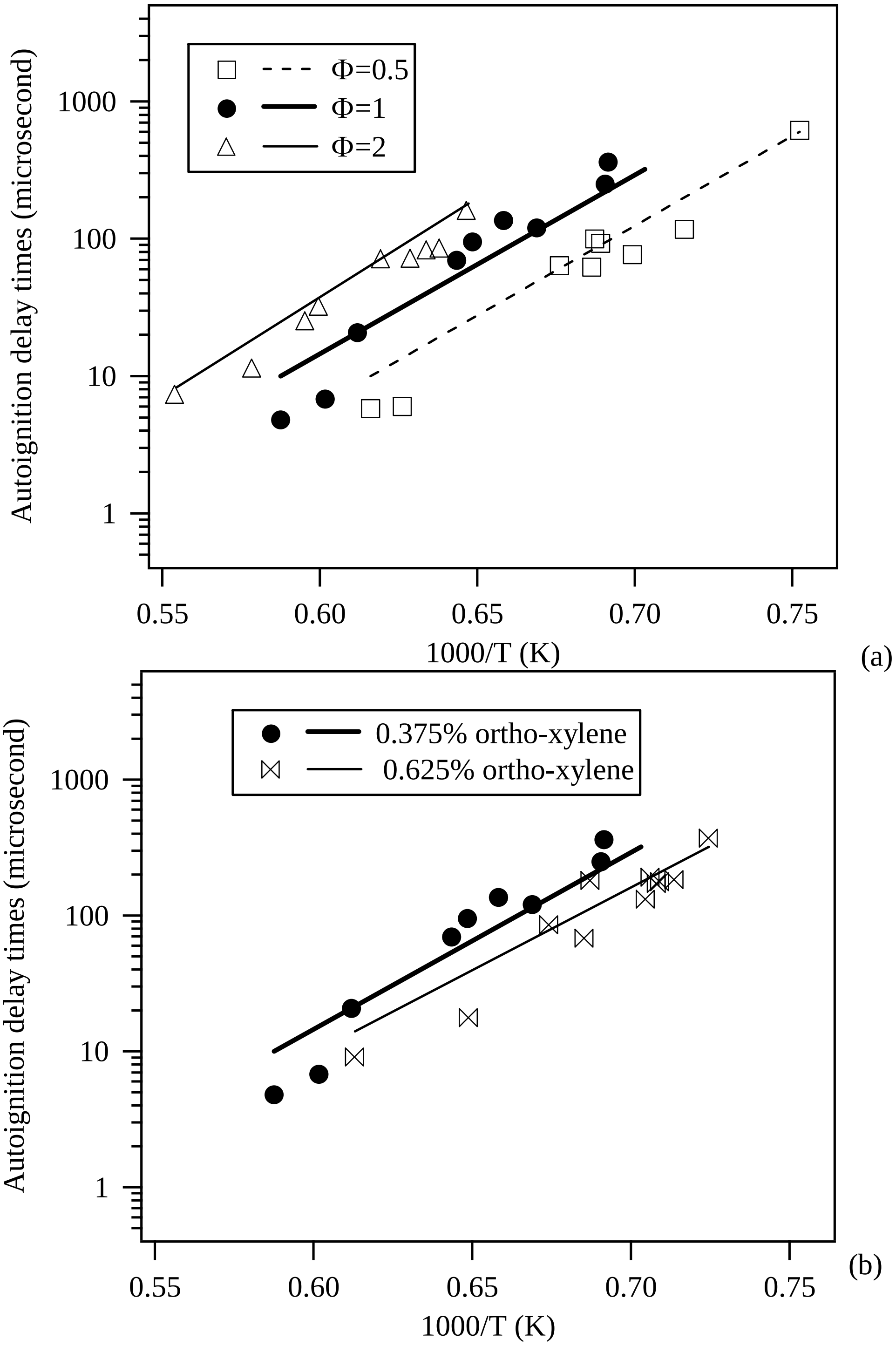

Figure 2

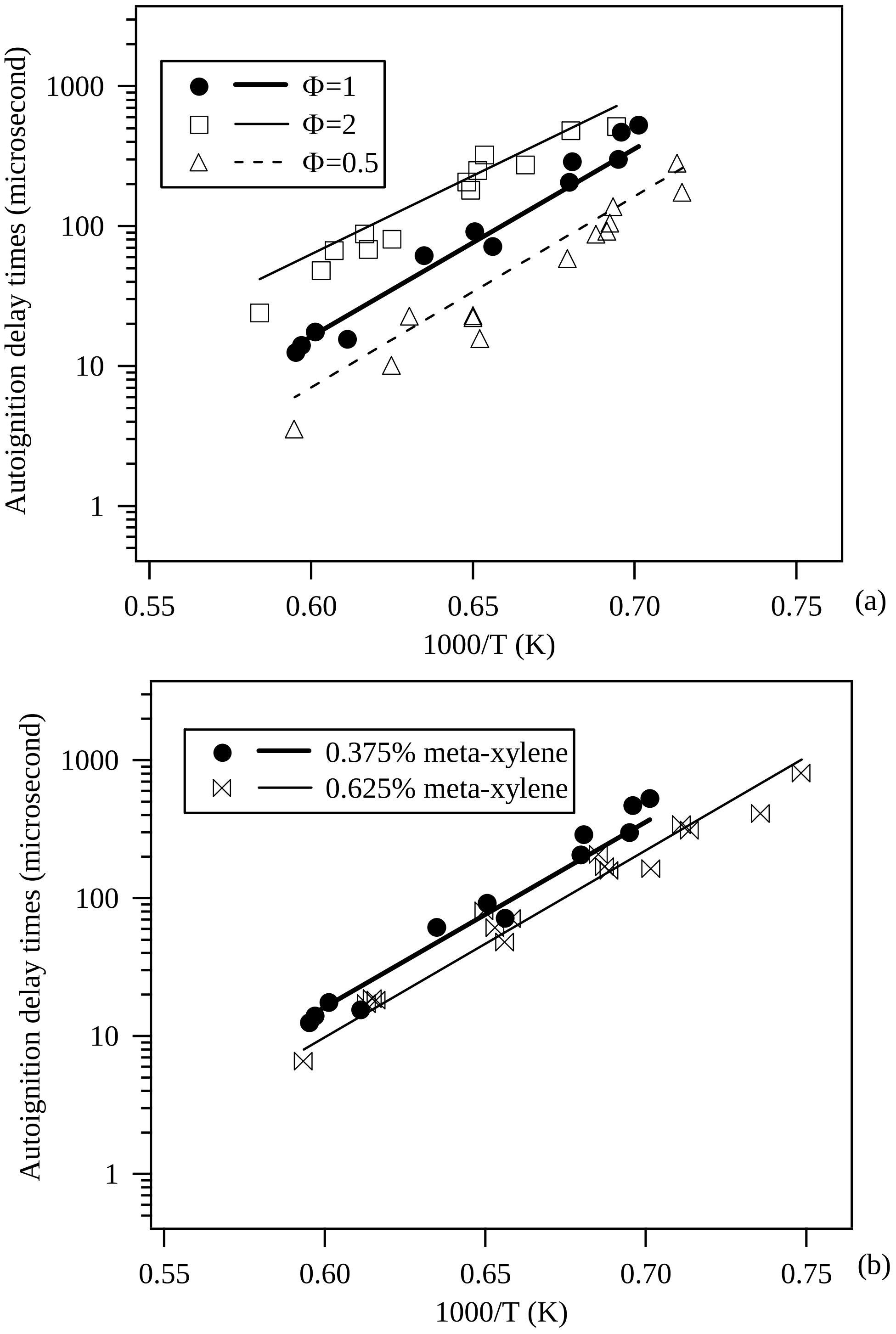

Figure 3

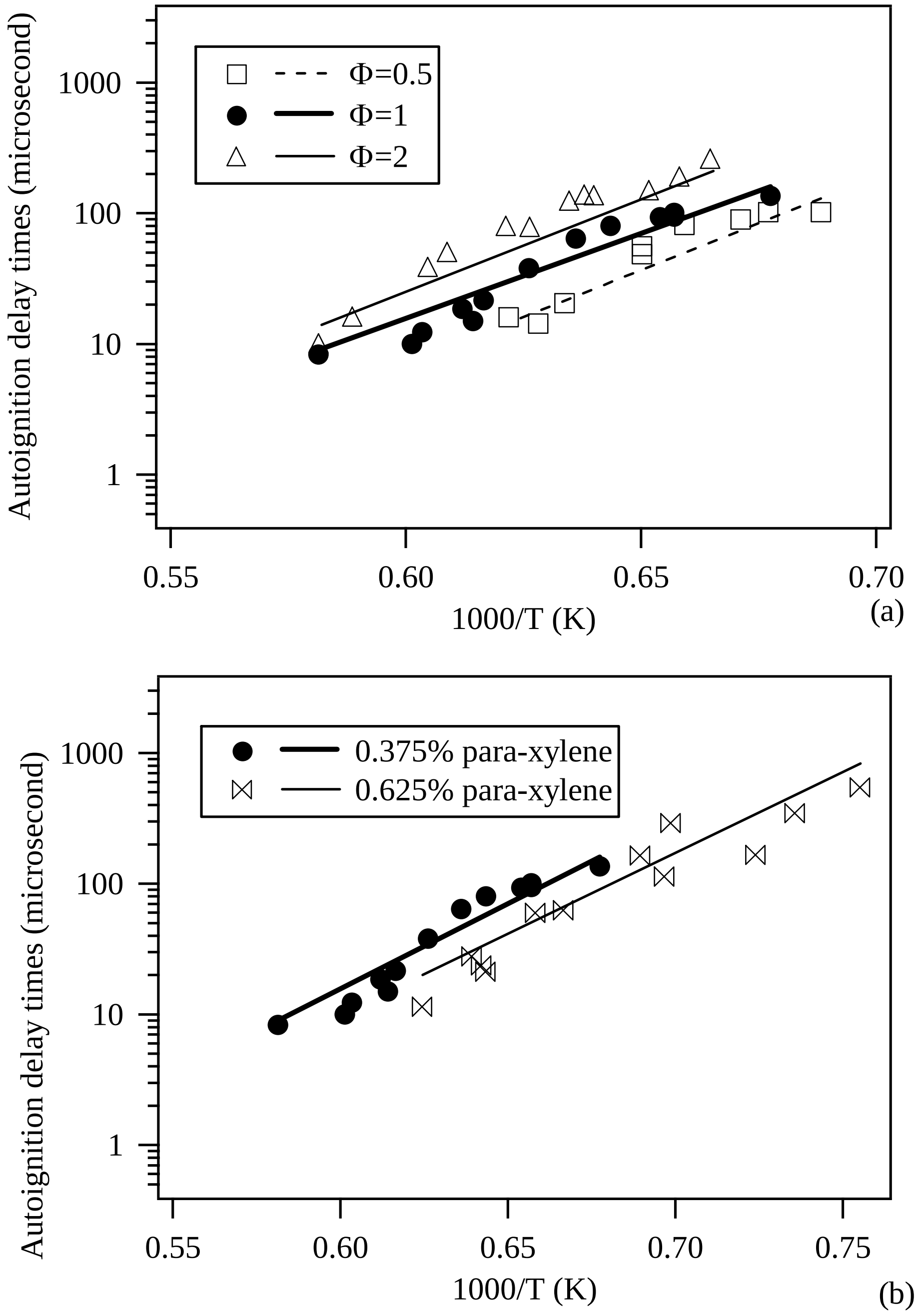

Figure 4

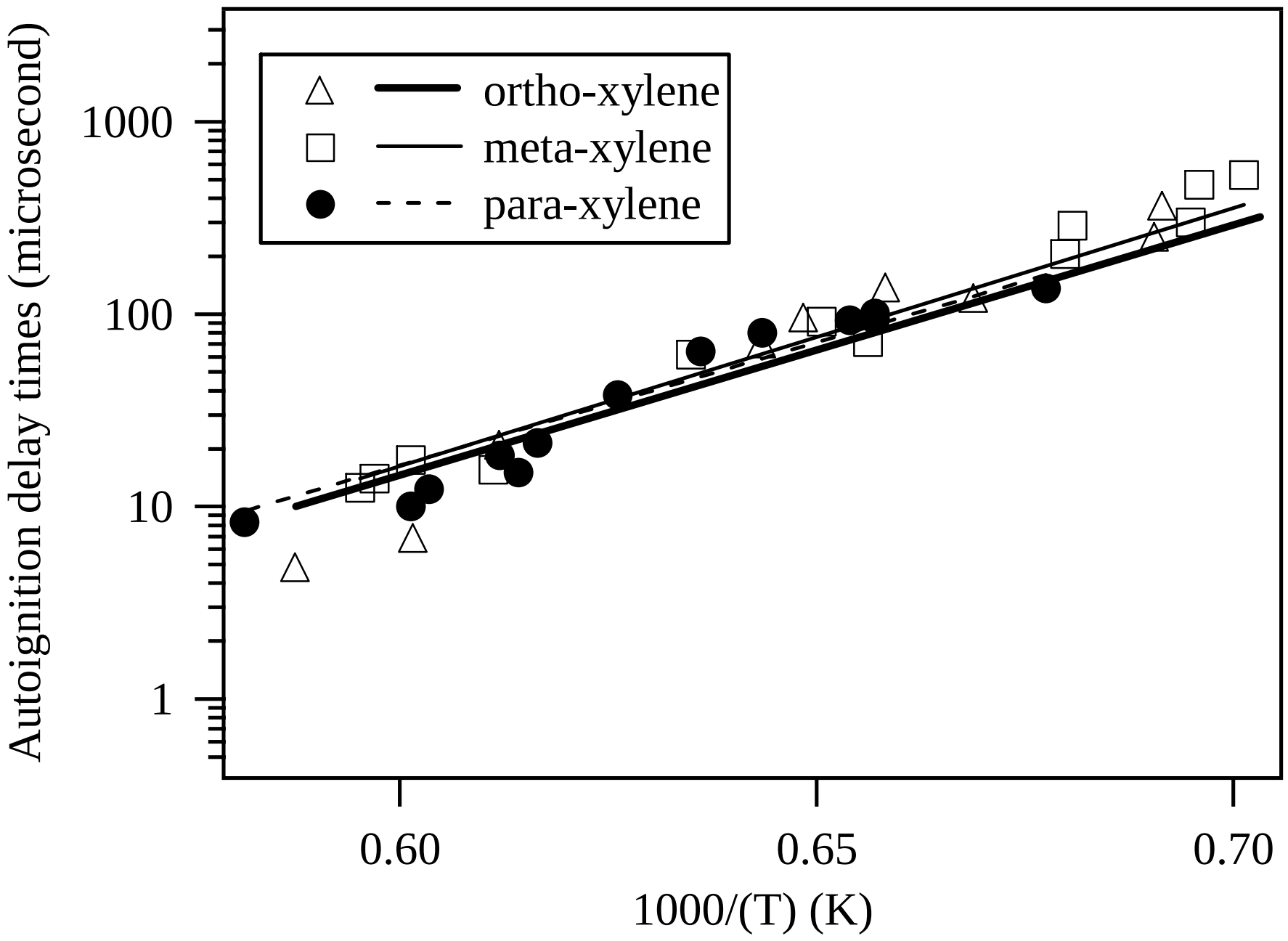

Figure 5

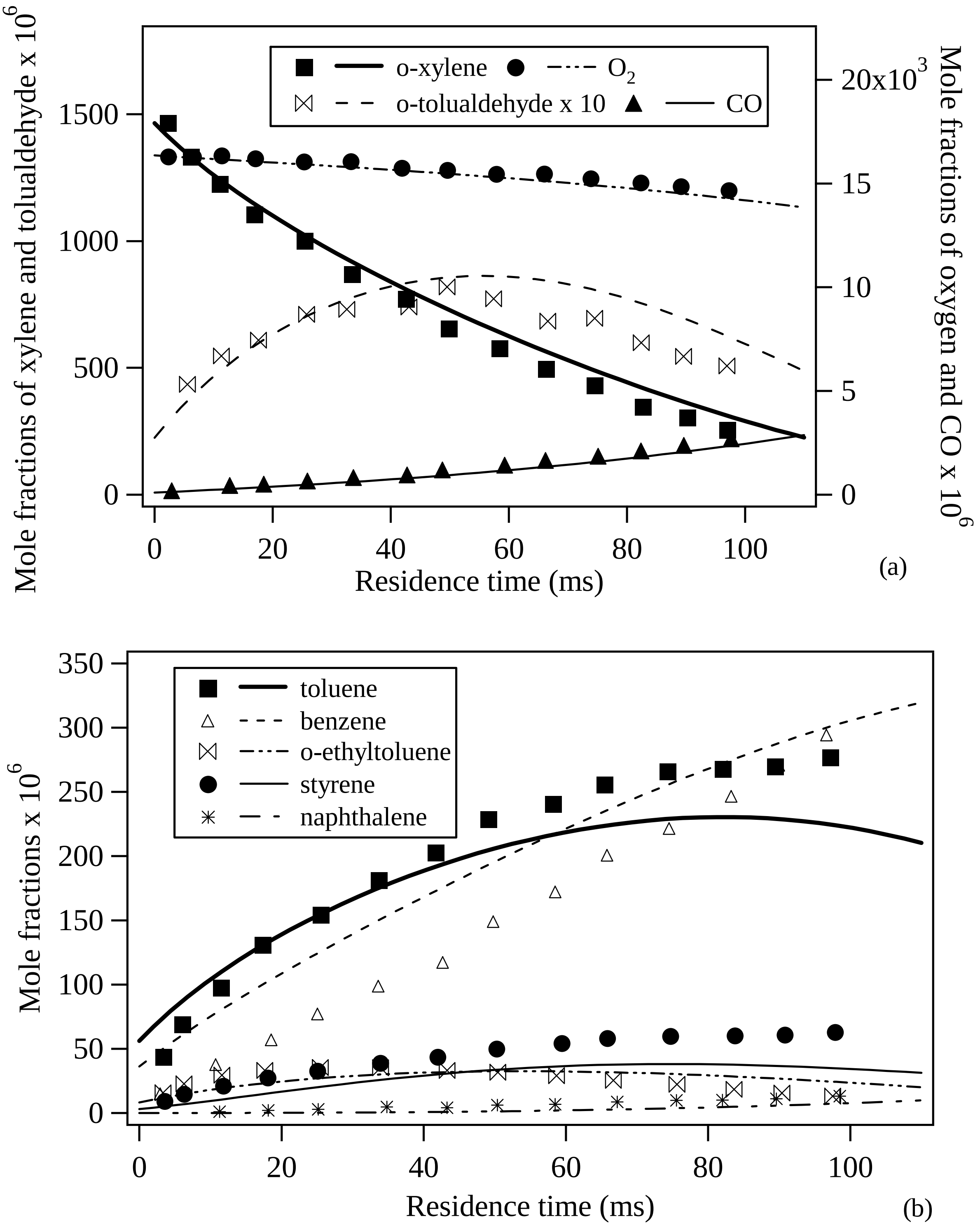

o-xylene
O2
o-tolualdehyde x 10
CO
Mole fractions of xylene and tolualdehyde x 10^6
Mole fractions of oxygen and CO x 10^6
20x10^3
Residence time (ms)
(a)
toluene
benzene
o-ethyltoluene
styrene
naphthalene
Mole fractions x 10^6
Residence time (ms)
(b)

Figure 6

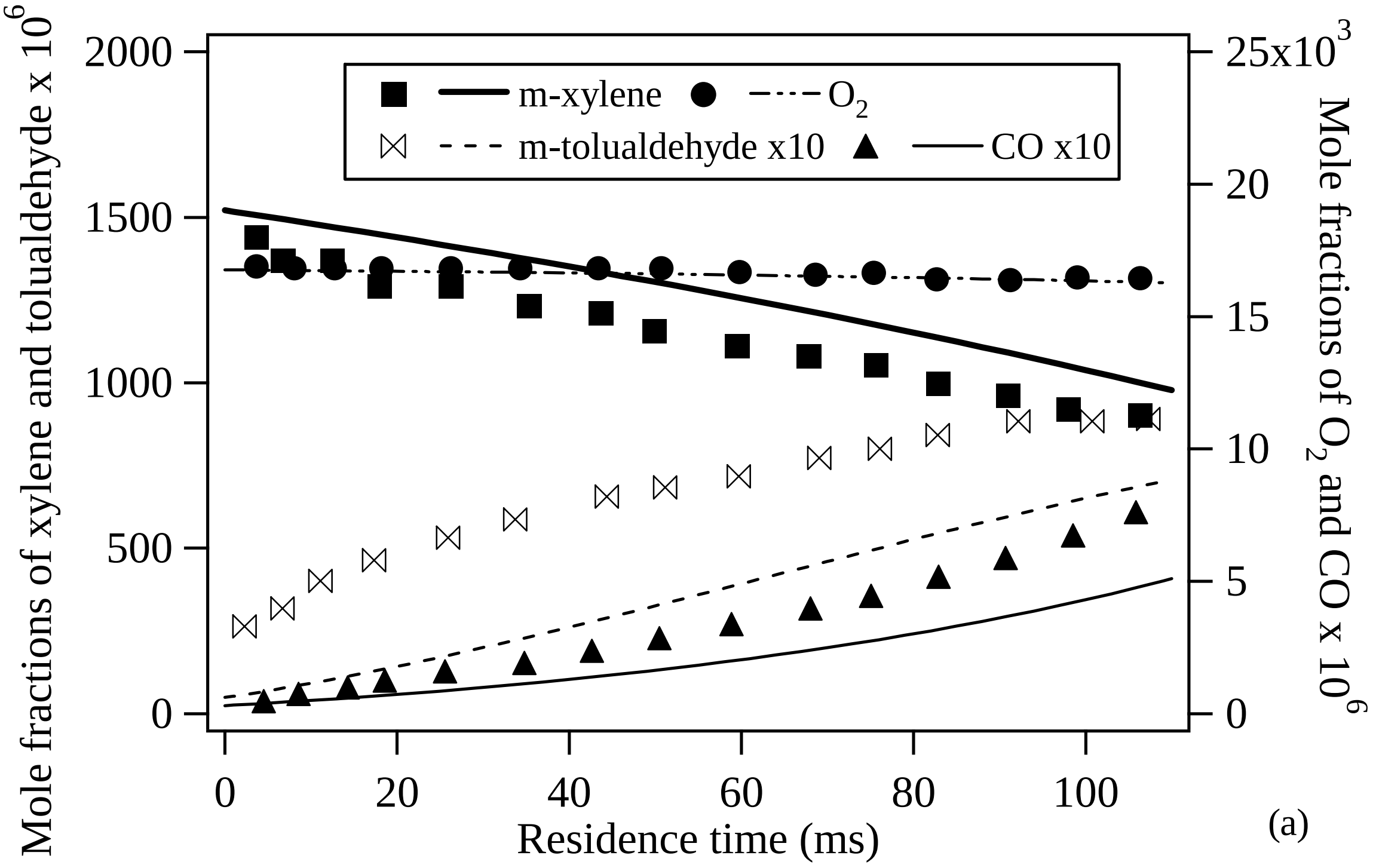

Mole fractions of xylene and tolualdehyde x 10^6
2000
1500
1000
500
0
m-xylene
O2
m-tolualdehyde x10
CO x10
25x10^3
20
15
10
5
0
Mole fractions of O2 and CO x 10^6
0
20
40
60
80
100
Residence time (ms)

(a)

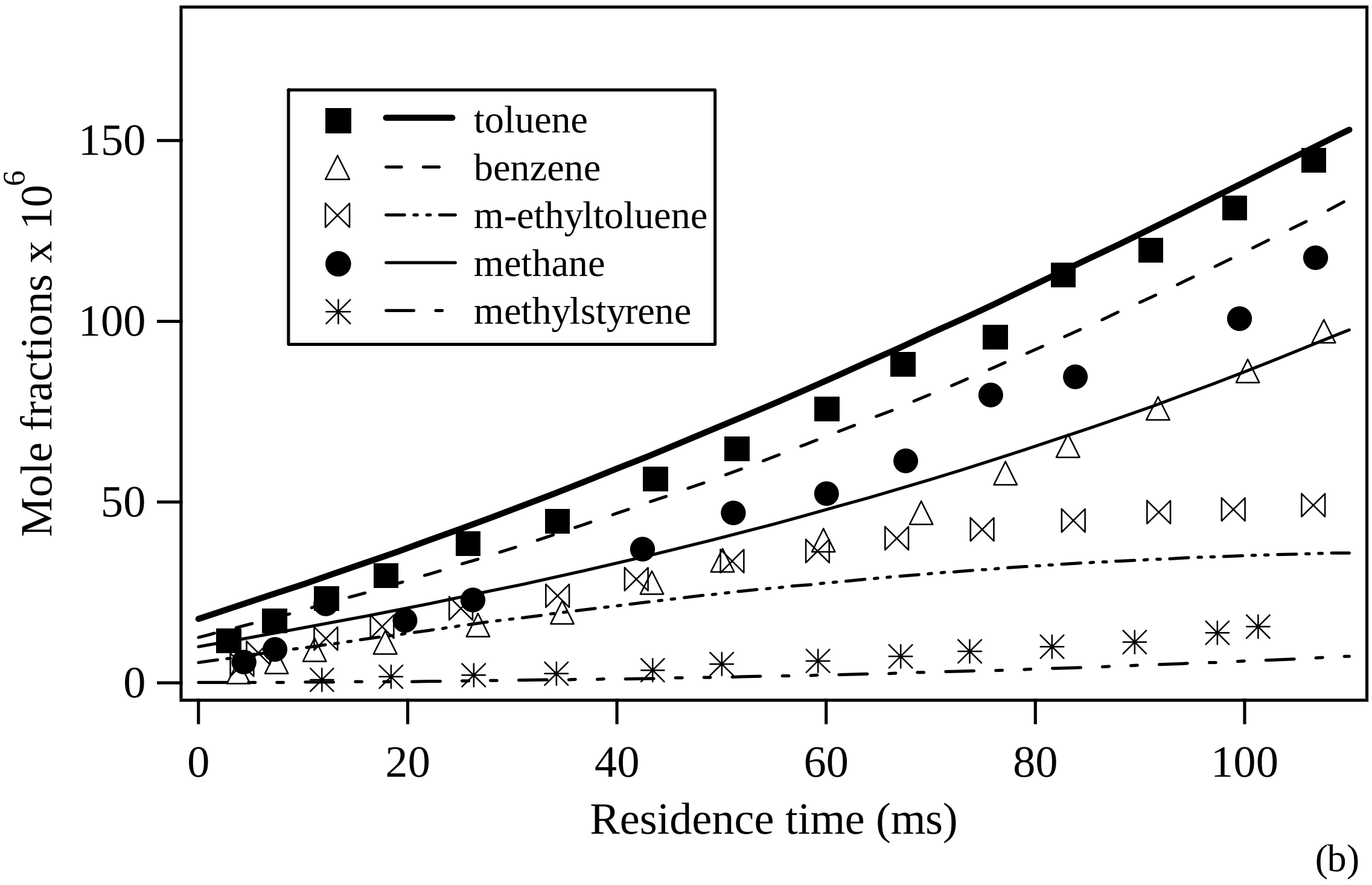

(b)

Figure 7

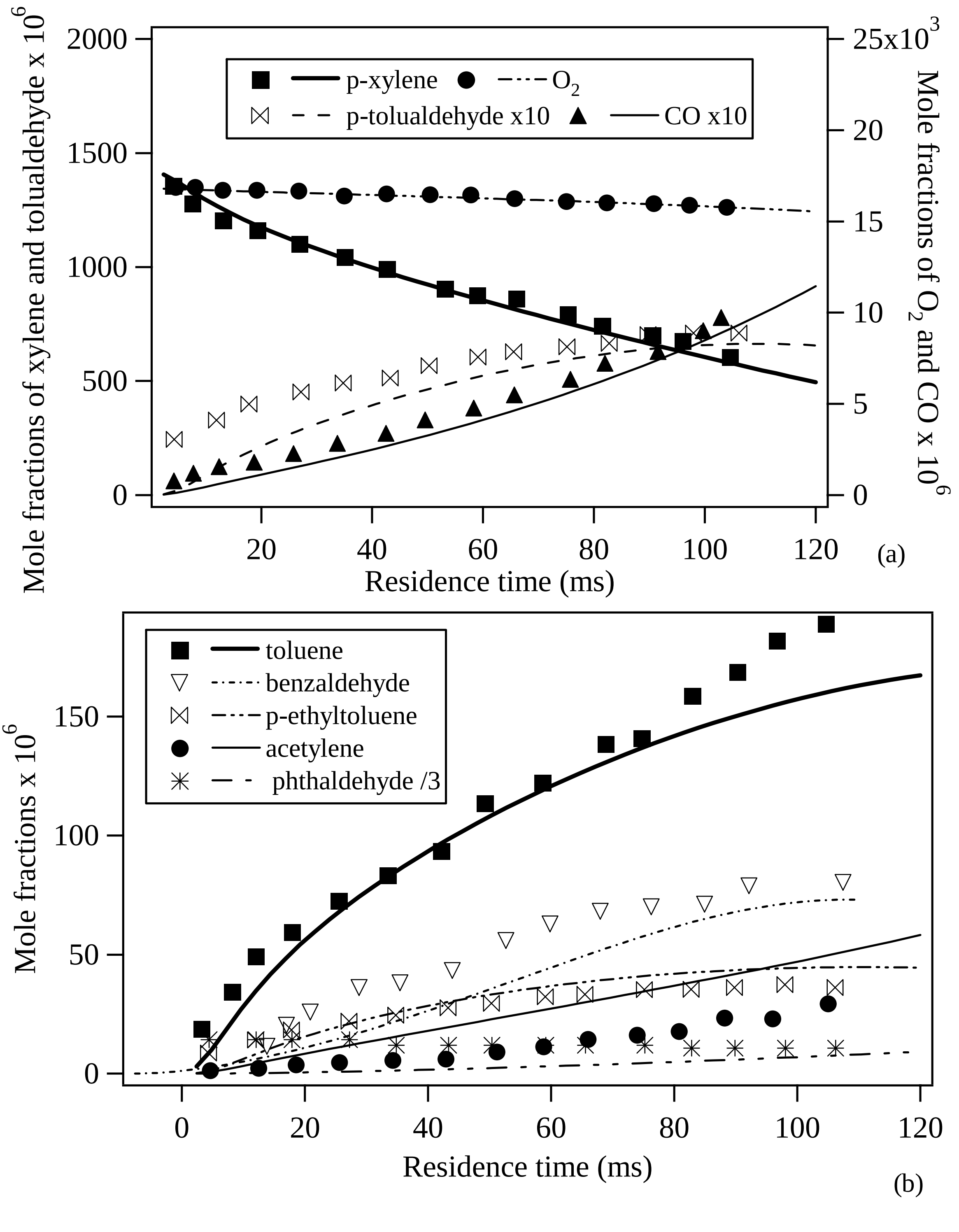

p-xylene
$O_2$
p-tolualdehyde x10
CO x10
Mole fractions of xylene and tolualdehyde x $10^6$
Mole fractions of $O_2$ and CO x $10^6$
2000
1500
1000
500
0
25x10[3]
20
15
10
5
0
20
40
60
80
100
120
Residence time (ms)
(a)
toluene
benzaldehyde
p-ethyltoluene
acetylene
phthaldehyde /3
Mole fractions x $10^6$
150
100
50
0
0
20
40
60
80
100
120
Residence time (ms)
(b)

Figure 8

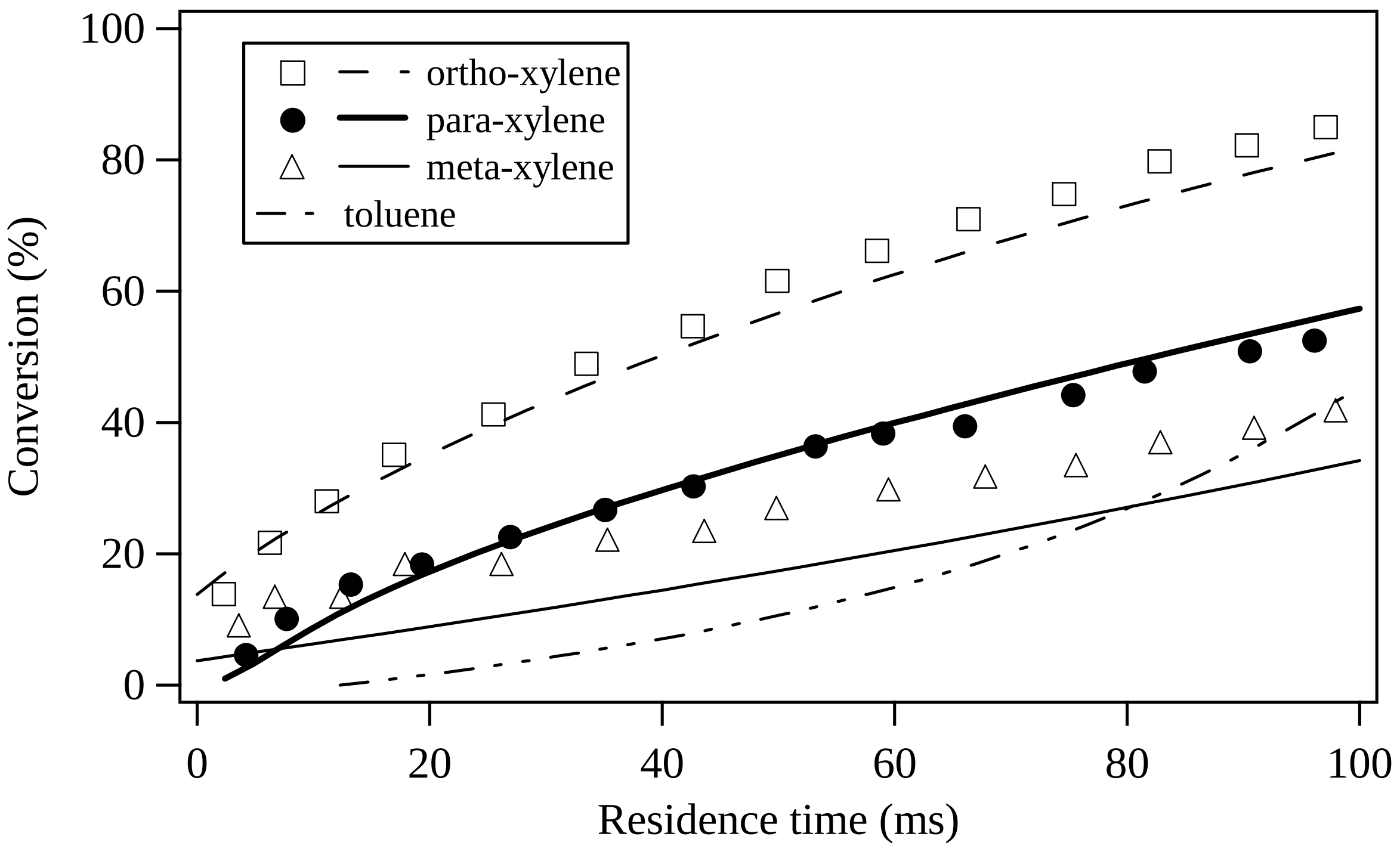

Figure 9

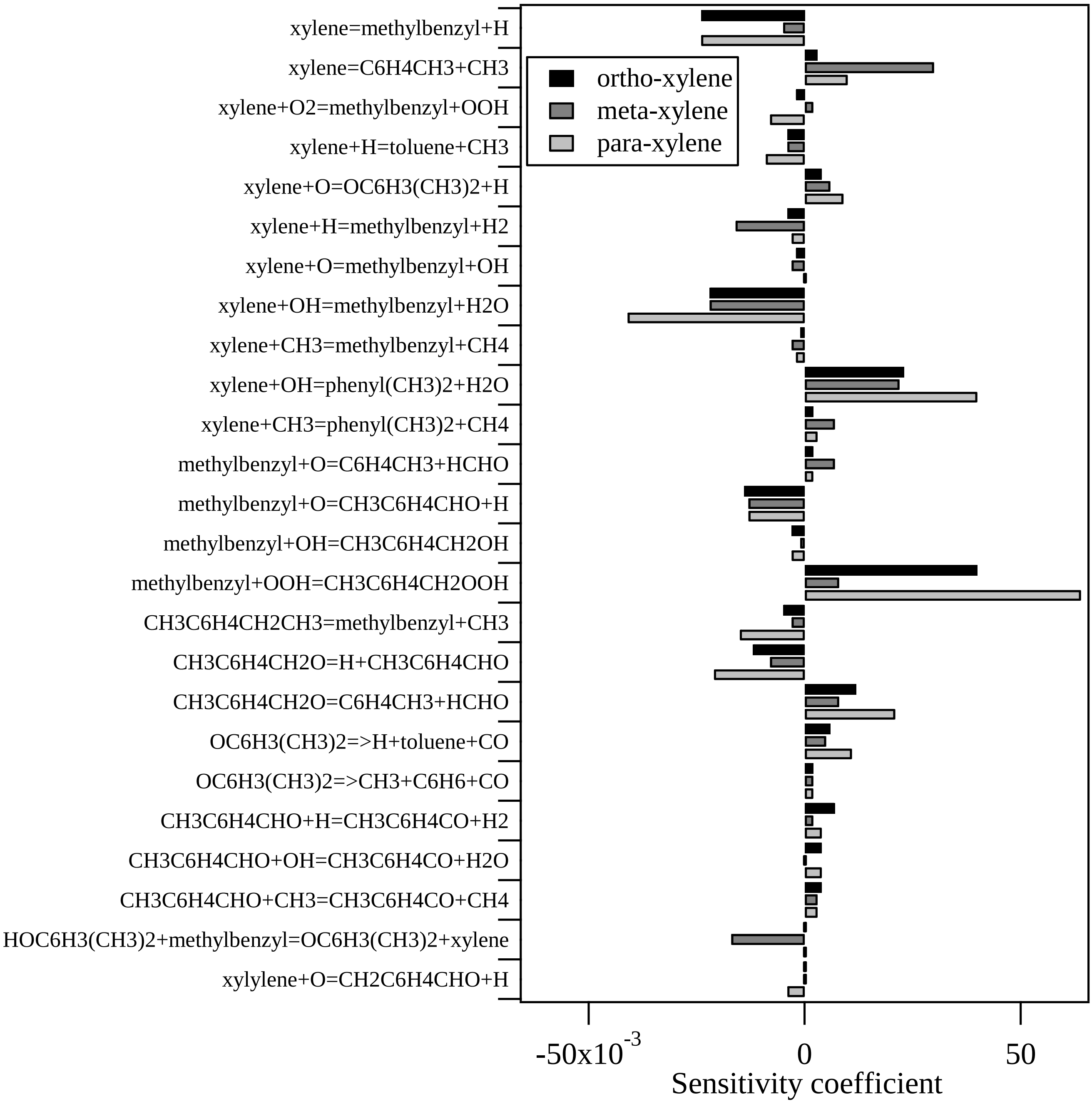

Figure 10

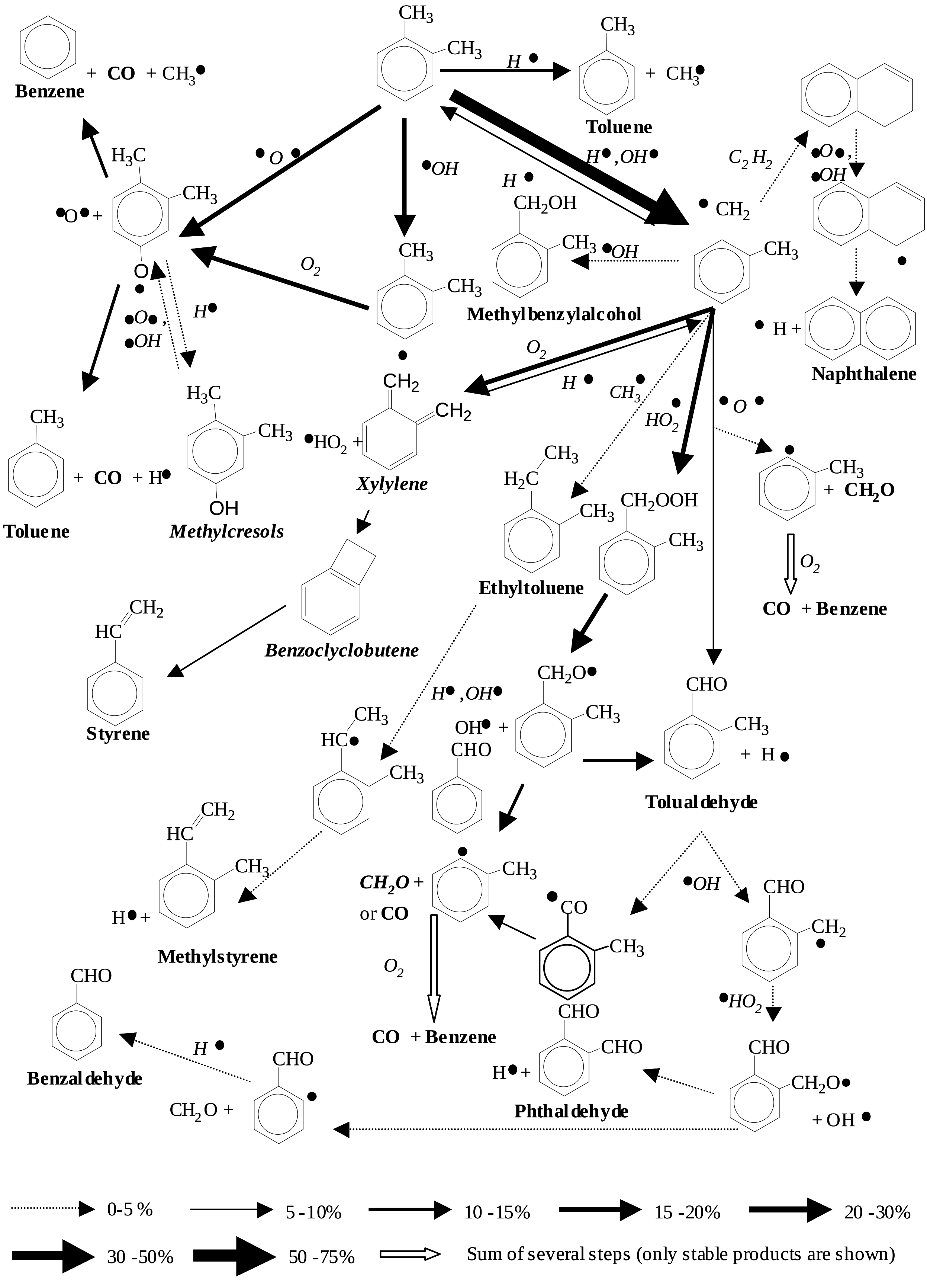
Benzene
+ CO + CH3•
H3C
CH3
•O• +
O
•O•
CH3
H•
Toluene
+ CH3•
•OH
H •
•O•,
•OH
H•
O2
CH2OH
CH3
•OH
Methylbenzylalcohol
• CH2
CH3
C2H2
•O•,
•OH
• H +
Naphthalene
•O•,
•OH
+ CO + H•
Toluene
H3C
CH3
OH
Methylcresols
CH2
CH2
•HO2 +
Xylylene
O2
H •
CH3•
HO2•
•O•
+ CH2O
O2
CO + Benzene
H2C
CH3
Ethyltoluene
CH2OOH
Benzoclyclobutene
HC
CH2
Styrene
H•,OH•
HC•
CH2O•
OH• +
CHO
CHO
+ H •
Tolualdehyde
HC
CH2
H•+
Methylstyrene
CH2O +
or CO
O2
CO + Benzene
•CO
•OH
CHO
CH2
•HO2
CHO
Benzaldehyde
H •
CH2O +
CHO
H•+
CHO
Phthaldehyde
CH2O•
+ OH •
0-5 %
5 -10%
10 -15%
15 -20%
20 -30%
30 -50%
50 -75%
Sum of several steps (only stable products are shown)

Figure 11

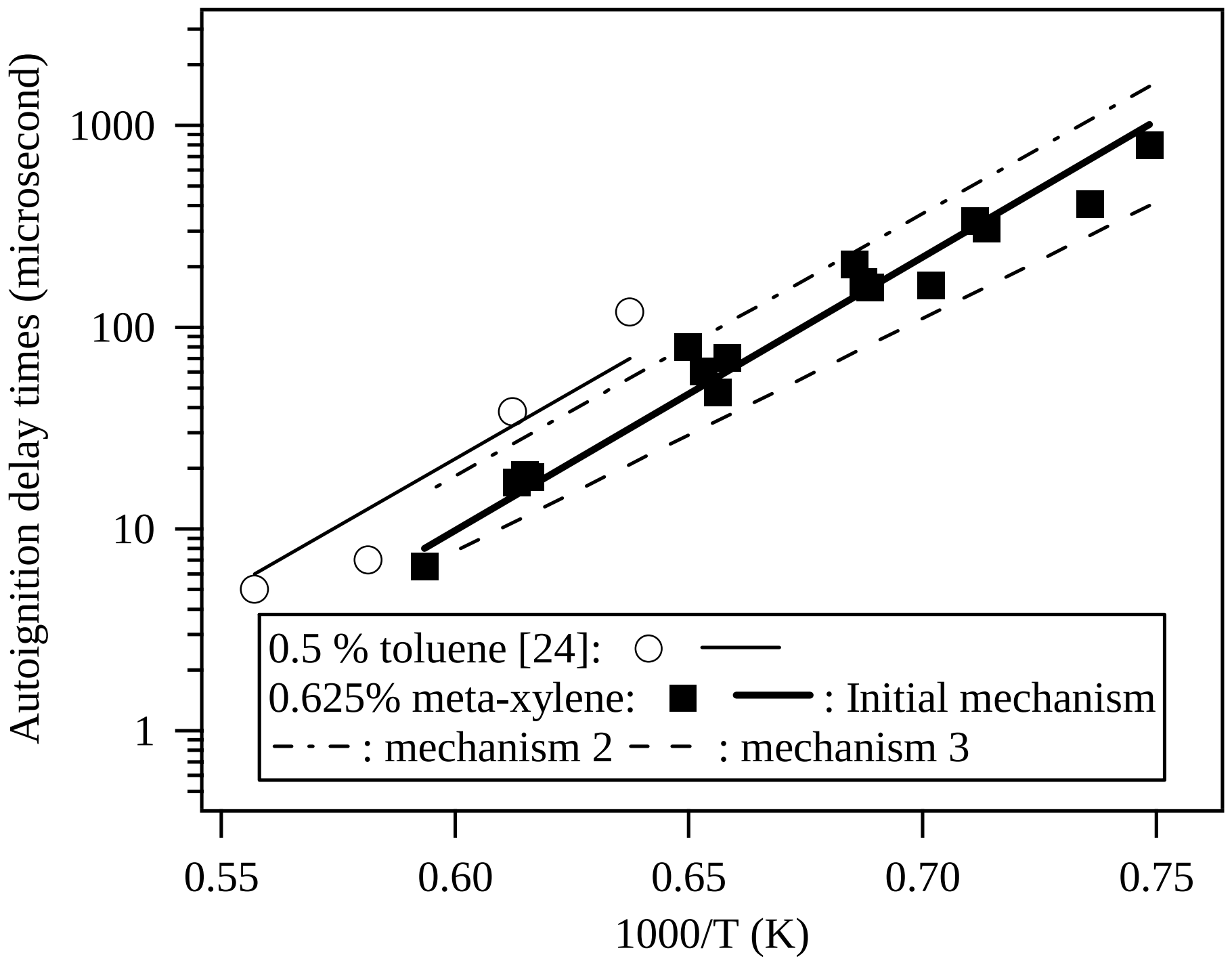

Figure 12

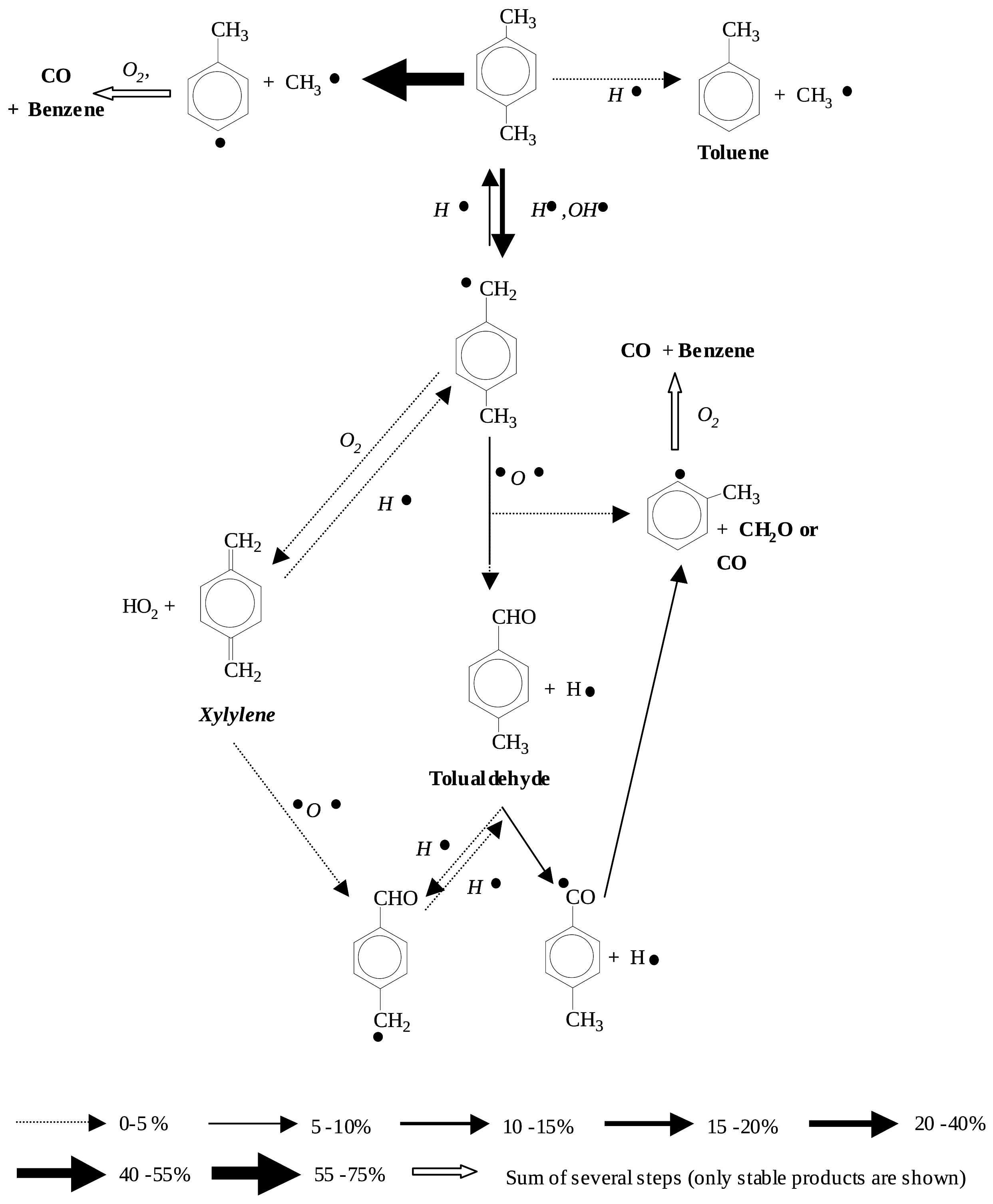
CH3
CO
O2,
+ Benzene
+ CH3
H
Toluene
H
H , OH
CH2
CO + Benzene
O2
O2
O
H
CH2
HO2 +
+ CH2O or
CO
CHO
+ H
Xylylene
Toluualdehyde
O
H
H
CHO
CO
+ H
CH2
CH3
0-5 %
5 -10%
10 -15%
15 -20%
20 -40%
40 -55%
55 -75%
Sum of several steps (only stable products are shown)